\documentclass[twocolumn]{aastex701}
\renewcommand{\today}{November 26, 2026}
\usepackage{amsmath}
\usepackage{needspace}
\usepackage{placeins}
\providecommand{\dodoi}[1]{doi:~\href{https://doi.org/#1}{\nolinkurl{#1}}}
\newcommand{\Fp}{F_+}
\newcommand{\Fm}{F_-}
\shorttitle{Why the eclipses switched off}
\shortauthors{Duan et al.}
\begin{document}

\title{Why the Eclipses Switched Off: Nodal Precession in a Misaligned Triple}

\author[0000-0002-6573-6719]{Xiao-Wei Duan}
\affiliation{State Key Laboratory of Dark Matter Physics, Tsung-Dao Lee Institute
\& School of Physics and Astronomy, Shanghai Jiao Tong University,
Shanghai 201210, China}
\email{duanxw@sjtu.edu.cn}
\author[0000-0001-6753-4611]{Guang-Yao Xiao}
\affiliation{State Key Laboratory of Dark Matter Physics, Tsung-Dao Lee Institute
\& School of Physics and Astronomy, Shanghai Jiao Tong University,
Shanghai 201210, China}
\email{gyxiao_tdli@sjtu.edu.cn}
\author[0000-0001-6039-0555]{Fabo Feng}
\affiliation{State Key Laboratory of Dark Matter Physics, Tsung-Dao Lee Institute
\& School of Physics and Astronomy, Shanghai Jiao Tong University,
Shanghai 201210, China}
\email{ffeng@sjtu.edu.cn}

\begin{abstract}
The mutual inclination of a hierarchical triple records how the system formed and sets how its inner orbit precesses. An inner binary that stops eclipsing has just left a narrow band of inclinations, so the end of its eclipses constrains the orbital geometry. TESS recorded eleven eclipses of a 12.35-day binary in Gaia DR3 3430035374498386432 (G3430) in 2021 and none in late 2023. LAMOST velocities of a second star, B, follow both this orbit and the 291-day orbit that Gaia also detects. The source is therefore a triple, A+(B+C), whose eclipsing pair is the astrometric companion of the brightest star, A. A single three-body model jointly fits the Gaia astrometric orbital solution, eclipse photometry, radial velocities, and spectral energy distribution. To our knowledge, this is the first such joint dynamical fit. It constrains the stellar masses, radii, effective temperatures and orbital geometry without stellar-evolution tracks or a mass--radius relation. The pair outweighs A but emits only about a fifth of the $G$-band light, so Gaia's photocenter orbit is about 60\% of A's. Two orientations of the inner node give mutual inclinations of 25$^\circ$ and 30$^\circ$, with undetermined relative probability. Nodal precession with a 49\ or 53\,yr period carried the inner orbit across the 6.1$^\circ$ eclipsing band in about two years, ending the eclipses. Assuming independent velocity errors, the two orientations predict that the eclipses return near 2039\ or 2049. Gaia astrometric orbits can thus find misaligned triples that eclipse surveys see only in brief seasons.
\end{abstract}

\section{Introduction} \label{sec:intro}
The mutual inclination of a hierarchical triple, the angle between its inner and outer orbits, records how the system formed and governs how it evolves \citep{Naoz2016,Tokovinin2017}. A misaligned inner orbit precesses about the total angular momentum \citep{Soderhjelm1975,Mazeh1976}, and strong misalignment drives Lidov--Kozai cycles, which with tidal friction can shrink the inner binary \citep{Fabrycky2007}. An inner binary eclipses only while its orbit is nearly edge-on, so a misaligned triple eclipses only in seasons that recur with the nodal cycle. Surveys that need eclipses over long baselines therefore favor aligned systems.

For compact triples, most mutual inclinations still come from eclipses, through eclipse timing \citep{Borkovits2016} and photodynamical models \citep{Carter2011,Borkovits2019,Czavalinga2023b}. Changing eclipse depths, and eclipses that appear and vanish, have revealed nodal precession in classical binaries \citep{Lacy1999,Torres2000,Zasche2012,Zasche2023} and in Kepler and TESS triples \citep{Borkovits2016,Borkovits2025,Borkovits2026}. Eclipse times have been fitted with velocities and the astrometry of a resolved outer orbit \citep{Akar2026}, and with Hipparcos and Gaia astrometry to characterize outer companions \citep{Xiao2026}. Astrometry combined with velocities measures mutual inclinations in planetary systems \citep{Xiao2026b}. Gaia orbits of unresolved triples have been compared with eclipse-timing orbits \citep{Czavalinga2023,Mitnyan2024} or combined with LAMOST spectra to identify triples \citep{He2025}. Gaia samples searched for dark companions also contain luminous inner pairs \citep{Xu2026,Tanikawa2026,ElBadry2026}.

In these studies the routes have remained separate. Photodynamical models can include velocities and broadband fluxes but have been compared with Gaia orbits only afterwards. Joint fits of astrometry and eclipse timing treat the eclipses as timing points and the outer orbit as Keplerian. Yet Gaia measures the orbit of the photocenter. The stellar fluxes that set the eclipse depths therefore also set its size, and one three-body state must reproduce the velocities, the eclipses and the Gaia orbit. To our knowledge, a Gaia astrometric orbital solution has not previously been fitted together with eclipse photometry in one dynamical model.

The triple Gaia DR3 3430035374498386432 (LAMOST J060032.43+265930.3, TIC~79317673, $G=11.76$; hereafter G3430) has a Gaia DR3 astrometric orbit of $293.7\pm3.9$\,d and double-lined spectra \citep{Kovalev2024}. \citet{Guo2025} reported its TESS eclipses in late 2021 and their absence two years later, and proposed that the source contains more than two stars. Our search for dark companions with velocities and astrometry also identified it as a probable triple \citep[][submitted]{Duan2026}. It has remained unclear which stars produce the two velocity signals and why the eclipses stopped. We show that the second star in the spectra, B, follows both the 12.35\,d eclipsing orbit and the 291\,d outer orbit. The eclipsing pair is therefore the astrometric companion of the brightest star, A, and the system is A+(B+C) (Figure~\ref{fig:archi}a). We fit one three-body model to the velocities, eclipse light curves, Gaia photocenter orbit and spectral energy distribution. The fit gives the mass and light of each star and the three-dimensional geometry, and shows that nodal precession carried the inner orbit out of eclipse in about two years.

\section{Observations} \label{sec:data}
\begin{figure*}
\centering\includegraphics[width=\textwidth]{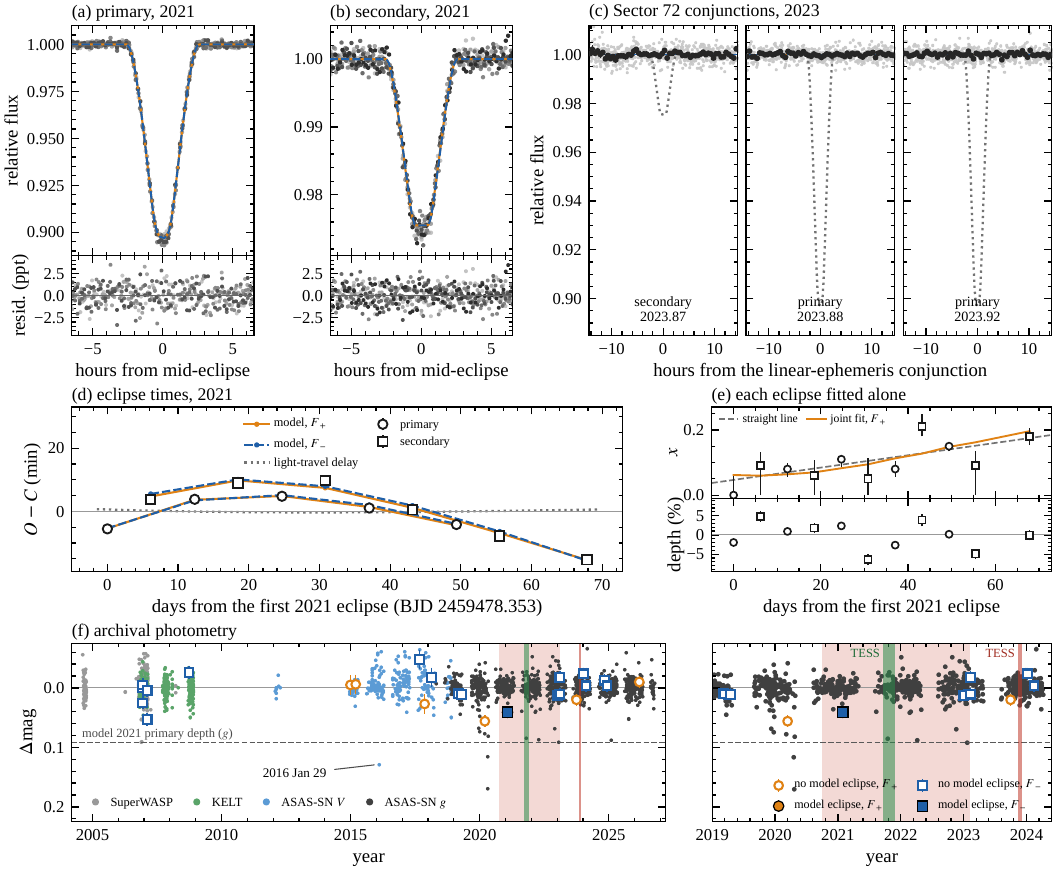}
\caption{Eclipses, eclipse times and the photometric record. (a, b) The 2021 primary and secondary eclipses, divided by their fitted baselines and aligned on their measured mid-times, with the model ($\Fp$ solid, $\Fm$ dashed) and the residuals. (c) The three Sector~72 windows of 2023, where neither median point predicts an eclipse, with the 2021 eclipse shape dotted. (d) Measured mid-times (open symbols) and model conjunction times, minus a reference ephemeris whose period, 12.34513\,d, is the slope of the five primary mid-times. Its epochs, ${\rm BJD_{TDB}}$ 2459478.3565\ for the primaries and 2459484.4726\ for the secondaries, refer the secondaries to phase 0.4954\ and remove their constant 81\,min offset from half phase. The 68\% intervals of the mid-times, from fitting each eclipse alone, are smaller than the symbols, and the mid-times are not a separate likelihood term of the joint fit. The dotted curve is the outer light-travel delay, less its linear trend. (e) Each eclipse fitted alone with the stars of the $\Fp$ median point (Appendix~\ref{app:impl}): the closest approach $x$ of B and C in units of $R_B+R_C$, with the best straight line (dashed) and the track of the joint fit, and the depth relative to the joint fit, with 68\% intervals. (f) Archival photometry, in full (left) and around 2021 (right). Shading marks the modeled season and vertical bands the TESS data. Large symbols are bins within an hour of a model primary conjunction (circles $\Fp$, squares $\Fm$), filled where the model eclipses. The dashed line is the model depth of the 2021 primary eclipses in the $g$ band, 92\,mmag (0.118\,mag in the TESS band). All model curves are drawn at the median points.\label{fig:ecl}}
\end{figure*}

The 10-minute TESS light curves of Sectors 43--45 \citep{Ricker2015,Caldwell2020} contain five primary and six secondary eclipses, 10.3\% and 2.6\% deep (Figure~\ref{fig:ecl}a, b). We fit the 1408\ fluxes within $\pm0.45$\,d of the eleven measured mid-times, not eclipse times, depths or durations. Sector~72 was observed at 200\,s cadence \citep{Huang2020} for 24.0\,d in late 2023. Its three windows around the conjunctions of the 2021 linear ephemeris contain 1547\ fluxes and no eclipse (Figure~\ref{fig:ecl}c). A search of the whole sector finds no dip deeper than 0.20\%, whereas eclipses 0.4\% deep would be recovered in at least 99\% of cases (Appendix~\ref{app:impl}).

\begin{figure*}
\centering\includegraphics[width=\textwidth]{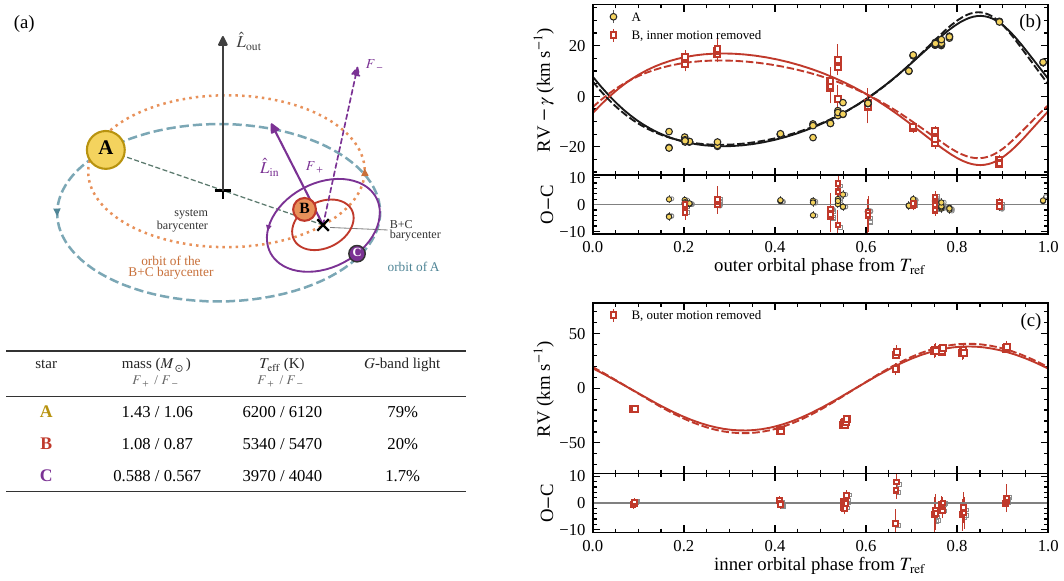}
\caption{Architecture of G3430 and the velocity evidence. (a) Schematic of the hierarchy A+(B+C) for the $\Fp$ median point (Section~\ref{sec:sampling}). The outer orbits of A and of the B+C barycenter ($P_{\rm out}=291$\,d) are drawn to scale, the inner orbits of B and C ($P_{\rm in}=12.35$\,d) are enlarged threefold, and the stellar radii are not to scale. The view shows both orbital planes, not the sky projection of Figure~\ref{fig:orbit}. Arrowheads show the direction of motion about the poles $\hat{L}_{\rm out}$ and $\hat{L}_{\rm in}$. The solid inner pole is that of $\Fp$ ($i_{\rm mut}=25^\circ$) and the dashed one that of $\Fm$ ($30^\circ$). The outer eccentricity is 0.26\ or 0.29, and the nodal period 49\ or 53\,yr. The table gives sampled medians of the masses and $T_{\rm eff}$ (Table~\ref{tab:full}) and the $G$-band light shares of the $\Fp$ median point. (b) Outer orbit: velocities of A (circles) and resolved velocities of B with its $\Fp$ inner motion removed (squares), which trace the B+C barycenter. (c) Inner orbit: the same velocities of B with the $\Fp$ barycentric motion removed. Error bars include the fitted jitter. Curves are the model at the median points ($\Fp$ solid, $\Fm$ dashed). The data are corrected with the $\Fp$ model, and the $\Fm$ curves show the same component of the $\Fm$ model, so the upper panels illustrate the components rather than the fit of $\Fm$. The lower panels give the residuals in km\,s$^{-1}$ of $\Fp$ (symbols as above) and of $\Fm$ (gray, offset in phase), each against its own model.\label{fig:archi}}
\end{figure*}

LAMOST \citep{Cui2012} medium-resolution spectra from 19\ nights in 2019.1--2023.1\ give 64\ velocities of A, the brightest star. Two-component broadening functions \citep{Rucinski2002} resolve a second star, B, in 28\ exposures. In the other 36\ exposures B is blended with A, and we take no velocity of B from them (Section~\ref{sec:sum}). A third star, C, is not detected in the spectra. B's velocity combines the motion of the B+C barycenter on the outer orbit, in antiphase with A, with B's own 12.35\,d motion within the pair (Figure~\ref{fig:archi}b, c). With the outer motion removed, B's resolved velocities alone recover the eclipsing period. The best orbit near the 1.39\,d that \citet{Kovalev2024} found for their primary component is worse by $\Delta\chi^2\geq394$. One star seen on both orbits thus fixes the architecture, A+(B+C).

From Gaia DR3 we take the Thiele--Innes constants, period, eccentricity and parallax of the orbital solution \citep{GaiaVallenari2023,Halbwachs2023,GaiaArenou2023} with their covariance. They correspond to a photocenter semimajor axis $a_0=1.13\pm0.08$\,mas and an inclination of $84.1\pm1.3^\circ$.

SuperWASP \citep{Pollacco2006,Butters2010}, KELT \citep{Pepper2007} and ASAS-SN \citep{Shappee2014,Kochanek2017} photometry from 2004.6\ to 2026.7\ enters as 2352\ bins of 0.04\,d (Figure~\ref{fig:ecl}f). The spectral energy distribution (SED) consists of ten magnitudes of the unresolved system: APASS $B$ and $V$ \citep{Henden2016}, Gaia $G$, $G_{\rm BP}$ and $G_{\rm RP}$, 2MASS $JHK_s$ \citep{Skrutskie2006} and WISE $W1$ and $W2$ \citep{Wright2010}.

\section{The Joint Three-body Model} \label{sec:model}
\subsection{Model and predicted data} \label{sec:trial}
Each trial set of parameters describes one complete system: the three masses, the parallax $\varpi$, six osculating Jacobi elements for each orbit, an effective temperature and a radius for each star, and a common metallicity and extinction. The elements refer to $T_{\rm ref}={\rm BJD_{TDB}}\ 2459477.352708$, one day before the first measured eclipse. Twelve limb-darkening, dilution and noise terms bring the total to 36\ sampled coordinates (Table~\ref{tab:full}).

One REBOUND integration \citep{Rein2012,Rein2015} from $T_{\rm ref}$ to the epochs of observation predicts all the data. The positions of B and C, corrected for the light-travel delay across the outer orbit, give the hidden fraction of each limb-darkened disk at every TESS time stamp. The same geometry thus sets both the shapes and the times of the eclipses, their absence in the 2023 windows and the dimming in each archival bin. At each Gaia transit predicted for the DR3 span the photocenter is the $G$-band flux-weighted mean position of the three stars,
\begin{equation}
\boldsymbol{r}_{\rm ph}(t)=\frac{\sum_{j}F_{G,j}\,\boldsymbol{r}_j(t)}{\sum_{j}F_{G,j}},\qquad j\in\{A,B,C\},
\label{eq:phot}
\end{equation}
including the small wobble of B and C about their barycenter. We fit its along-scan coordinates with Gaia's model and compare the resulting orbital solution with the catalog. The comparison thus includes the precession of the outer orbit between the Gaia epochs and $T_{\rm ref}$. Each star's magnitude in each band follows from its temperature, radius and mass through the MIST bolometric-correction tables \citep{Choi2016,Dotter2016}, with extinction \citep{Cardelli1989}, and the summed fluxes give the SED. The same tables give the flux ratios for the eclipses and the photocenter, so the eclipse dilution, the photocenter weights and the SED cannot be adjusted separately.

An orbit's inclination is the angle between its angular momentum and the direction to the observer, so $i<90^\circ$ is counterclockwise motion on the sky. The longitude of the node, $\Omega$, is the position angle, north through east, of the receding node. The mutual inclination is the angle between the two angular momenta,
\begin{equation}
\cos i_{\rm mut}=\cos i_{\rm in}\cos i_{\rm out}+\sin i_{\rm in}\sin i_{\rm out}\cos(\Omega_{\rm in}-\Omega_{\rm out}),
\label{eq:imut}
\end{equation}
evaluated, like all elements quoted below, at $T_{\rm ref}$.

\subsection{What each data set constrains} \label{sec:like}
We sample the masses with uniform priors and impose no isochrone or mass--radius relation. The velocity amplitudes of A and of the B+C barycenter, with the period, eccentricity and inclination of the outer orbit, give $M_A$ and $M_B+M_C$. B's motion within the pair, with the inner inclination from the eclipses, separates $M_B$ from $M_C$. Eclipse depths measure the surface-brightness ratio of B and C and the dilution by A. Gaia's orbit, compared with A's orbit from the velocities and parallax, measures the pair's share of the $G$-band light, and the SED the total flux and its color.

\begin{figure*}
\centering\includegraphics[width=\textwidth]{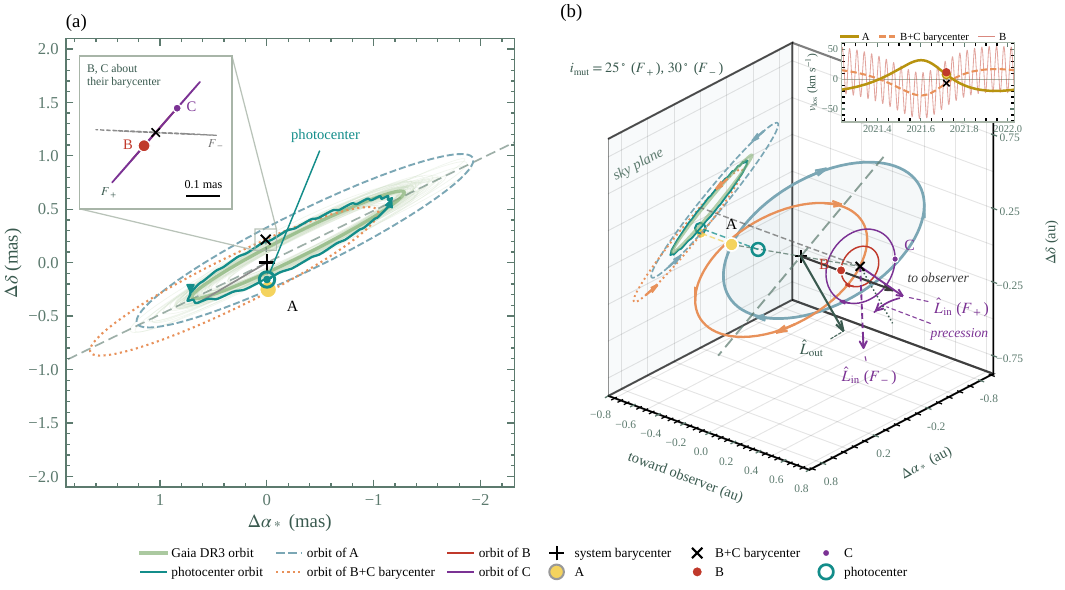}
\caption{Orbit geometry of the $\Fp$ median point over one outer period of the three-body integration from $T_{\rm ref}$ (2021.72). (a) Sky plane. The Gaia DR3 catalog orbit (green, with thin covariance draws) is an effective Keplerian solution for the observations of 2014.7--2017.3. The model photocenter in 2021.72\ (teal) moves counterclockwise on the sky (arrowheads). Dashed and dotted curves are the orbits of A and of the B+C barycenter about the system barycenter. The fit compares the model with the catalog at the Gaia epochs (Section~\ref{sec:trial}). Between those epochs and $T_{\rm ref}$ the outer inclination of this point changes by $-1.9^\circ$. The ripple of the photocenter is the motion of B and C. The inset shows B and C about their barycenter at true scale, with the $\Fm$ inner orbit gray dashed. (b) The same configuration in three dimensions, with the inner orbits enlarged threefold. The black arrow points to the observer. Solid arrows are the angular momenta of $\Fp$, which point away from the reader, so the motion appears clockwise. The dashed arrow is $\hat{L}_{\rm in}$ of $\Fm$, placed relative to the $\Fp$ outer orbit with the components it has in the frame of its own outer orbit (pole $\hat{L}_{\rm out}$, first axis toward the projected line of sight). The dotted line through the B+C barycenter is parallel to $\hat{L}_{\rm out}$, and each inner pole is inclined to it by $i_{\rm mut}$ (sampled medians). The curved arrow shows the direction in which $\hat{L}_{\rm in}$ precesses about the total angular momentum, $2.6^\circ$ from $\hat{L}_{\rm out}$ (Section~\ref{sec:cone}). The inset gives the line-of-sight velocities of A, the B+C barycenter and B over the outer orbit that contains $T_{\rm ref}$, 2021.23--2022.03. Apart from the Gaia orbit and its draws, curves are the model at the median points.\label{fig:orbit}}
\end{figure*}

Gaia fixes $i_{\rm out}$ and $\Omega_{\rm out}$, and the eclipse shapes fix $i_{\rm in}$, but nothing measures $\Omega_{\rm in}$ directly. Only the dynamics constrain it: the curvature of the eclipse times over 68\,d (Figure~\ref{fig:ecl}d), the drift of the eclipse geometry within 2021, and the absence of eclipses in the 2023 windows. Because both orbits are nearly edge-on, $\cos i_{\rm mut}\approx\cos(\Omega_{\rm in}-\Omega_{\rm out})$, and either sign of the node difference can fit. We label the two branches $F_\pm$ by the sign of $\sin(\Omega_{\rm in}-\Omega_{\rm out})$. Seen from Earth, $\hat{L}_{\rm in}$ lies on opposite sides of $\hat{L}_{\rm out}$ in the two branches (Figure~\ref{fig:orbit}b).

\subsection{Sampling and fitted points} \label{sec:sampling}
Each branch was sampled with a parallel-tempered ensemble sampler, following \citet{Vousden2016}, in three independent runs of 80{,}000\ steps (Appendix~\ref{app:diag}). Quoted values are medians and central 68\% intervals of the three runs of a branch, pooled with equal weight, and are conditional on the model. Table~\ref{tab:sys} gives their precision and their shifts under other data treatments. Each run stays on one branch, so the runs do not determine the relative probability of $\Fp$ and $\Fm$. Where two values are quoted without labels, the first refers to $\Fp$.

\begin{figure*}
\centering\includegraphics[width=\textwidth]{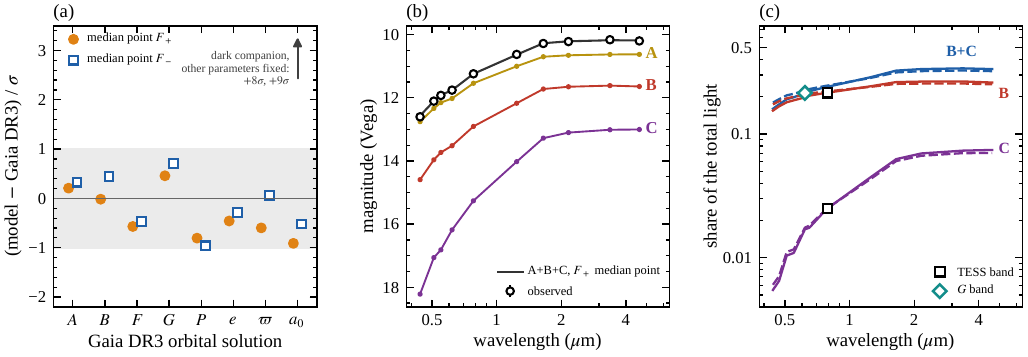}
\caption{One set of stars for the photocenter, the SED and the eclipse depths. (a) Model minus catalog, in units of the catalog error, for the seven fitted quantities of the Gaia DR3 orbital solution (Thiele--Innes constants $A$, $B$, $F$, $G$, period, eccentricity and parallax) and for the photocenter semimajor axis $a_0$, which is derived from them. Because the seven quantities are correlated, their $\chi^2$ in Section~\ref{sec:fit} uses the full covariance. The arrow marks where a dark companion of the same mass would leave the photocenter with all other parameters held fixed. This is not the significance of a fitted dark-companion model. (b) The SED with the contributions of A, B and C. Lines join the model magnitudes of the ten bands. (c) Shares of the total light ($\Fp$ solid, $\Fm$ dashed). The TESS-band shares set the eclipse depths, and the $G$-band share of the pair sets the photocenter. Model values are those of the median points.\label{fig:closure}}
\end{figure*}

Model curves in the figures are drawn at single sets of parameters, the fitted points, and carry no uncertainty band. The \textit{median point} of a branch, the vector of marginal medians of its sampled coordinates, is our representative solution. It need not maximize the posterior, and its curves are not pointwise medians of the sampled predictions. \textit{Profile points} hold the node difference at the value that gives a chosen percentile of $i_{\rm mut}$ and re-optimize the other coordinates. The fitted points are the median point and five profile points of each branch. Season dates and forecasts are medians and 68\% intervals from integrations of 2304\ sampled states of each branch (Appendix~\ref{app:points} and Table~\ref{tab:sampled}).

\section{Results} \label{sec:results}
\subsection{Fit to the data} \label{sec:fit}
At the median point of either branch the model reproduces every data set (Figures~\ref{fig:archi}--\ref{fig:closure}). The 1408\ fluxes of the 2021 TESS windows give $\chi^2=1360$ and 1367. The measured mid-times depart from a linear ephemeris by 25\,min peak to peak. The model conjunctions follow them to 0.9\ and 1.1\,min rms. The timing enters the fit through the fluxes at their observed times, and the measured mid-times are not a separate likelihood term. Velocity residuals have an rms of 1.6\ and 1.7\,km\,s$^{-1}$ for A and 3.0\ and 3.5\,km\,s$^{-1}$ for the resolved measurements of B. The seven Gaia quantities are matched with $\chi^2=2.1$ and 3.4, none deviating by more than 1.0$\sigma$, and the ten magnitudes within 0.8$\sigma$. Neither median point produces an eclipse in the Sector~72 windows.

\subsection{Masses and the luminous companion} \label{sec:masses}\label{sec:light}
The dynamical masses are $M_A=1.429^{+0.271}_{-0.230}\,M_\odot$, $M_B=1.080^{+0.148}_{-0.131}\,M_\odot$ and $M_C=0.588^{+0.015}_{-0.015}\,M_\odot$ for $\Fp$, and $M_A=1.062^{+0.173}_{-0.152}\,M_\odot$, $M_B=0.873^{+0.104}_{-0.095}\,M_\odot$ and $M_C=0.567^{+0.016}_{-0.016}\,M_\odot$ for $\Fm$ (Table~\ref{tab:full}). The pair has $1.17^{+0.11}_{-0.10}$ or $1.36^{+0.12}_{-0.10}$ times A's mass but emits only about a fifth of the $G$-band light, $21.4^{+0.7}_{-0.7}\%$ or $22.4^{+0.8}_{-0.8}\%$ (Figure~\ref{fig:closure}c). One main-sequence star of that mass would outshine A, whereas two less massive stars do not \citep{Shahaf2019}. Light from B+C pulls the photocenter toward the barycenter, to a fraction $\xi=0.60^{+0.01}_{-0.01}$ or $0.61^{+0.01}_{-0.01}$ of A's displacement. At the $\Fp$ median point, A's orbit has a semimajor axis of 1.76\,mas and the predicted photocenter orbit 1.06\,mas, close to Gaia's $1.13\pm0.08$\,mas (Figures~\ref{fig:orbit}a and \ref{fig:closure}a). With all other parameters fixed, a dark companion of the same mass would leave the photocenter on A, 8$\sigma$ or 9$\sigma$ from the Gaia value. This offset is not the significance of a fitted dark-companion model. The astrometric companion is therefore the eclipsing pair, and the same three stars reproduce the SED and the eclipse depths (Figure~\ref{fig:closure}b, c).

Unless stated otherwise, quoted values come from the reference fit, which assumes independent velocity errors. Allowing a common velocity offset for each night, which the data prefer, lowers the three masses, $M_A$ to $1.253^{+0.276}_{-0.233}$ or $0.846^{+0.173}_{-0.145}\,M_\odot$, by 0.7\ and 1.3\ times the half-width of its 68\% interval in the reference fit (Table~\ref{tab:sys}). We report the fit with these offsets as a sensitivity test. Its incomplete mixing limits the precision of this comparison (Appendix~\ref{app:impl}). The intervals of the reference fit do not include the shifts between the two noise models.

No isochrone or stellar-evolution mass--radius--age relation is imposed in the fit, although its light ratios use the MIST bolometric corrections (Section~\ref{sec:trial}). As a check outside the fit, we weighted the sampled states by how well their radii and temperatures match MIST isochrones of a single age \citep[][Appendix~\ref{app:ms}]{Dotter2026}. The isochrones agree with the $\Fm$ masses and with the lower part of the $\Fp$ intervals. Weighting by them gives $M_A=1.14^{+0.07}_{-0.08}$ and $M_B=0.91^{+0.04}_{-0.05}\,M_\odot$ on $\Fp$, and $1.08^{+0.07}_{-0.07}$ and $0.88^{+0.04}_{-0.04}\,M_\odot$ on $\Fm$. The mean weight of $\Fp$ is 3.5\ times smaller. This consistency factor is not a ratio of branch probabilities, and we do not use it to weight the branches.

\subsection{Geometry and the two node branches} \label{sec:geom}
The mutual inclination at $T_{\rm ref}$ is
\begin{equation}
i_{\rm mut}=24.72^{+2.47}_{-2.09}\ {\rm deg}\ (\Fp),\qquad 29.54^{+3.46}_{-3.04}\ {\rm deg}\ (\Fm).
\end{equation}
\begin{figure*}
\centering\includegraphics[width=\textwidth]{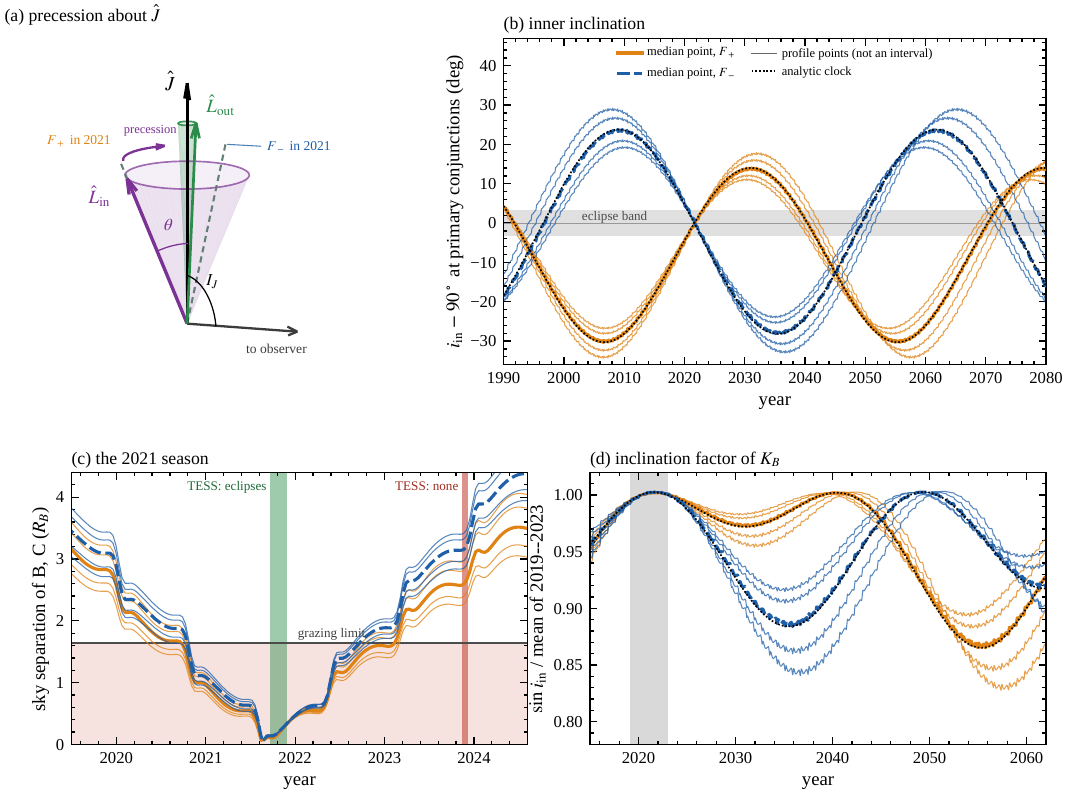}
\caption{Why the eclipses switched off. (a) Schematic of the precession about the total angular momentum $\hat{J}$, with the angles of the $\Fp$ median point. $\hat{L}_{\rm in}$ moves on a cone of half-angle $\theta$ and $\hat{L}_{\rm out}$ on a small cone opposite. The pair eclipses only near the two dashed orientations, where $\hat{L}_{\rm in}$ is perpendicular to the line of sight. $\Fp$ sits at one of them in 2021, and $\Fm$ at the other on its own, similar cone. (b) Inner inclination at primary conjunctions from REBOUND for the median points (thick; $\Fp$ solid, $\Fm$ dashed) and the profile points (thin), with the clock of Equation~(\ref{eq:clock}) (dotted). The profile points are fitted configurations at the 5th to 95th percentiles of $i_{\rm mut}$, not a sampled interval (Table~\ref{tab:sampled}). The gray band is Equation~(\ref{eq:band}), the limit for a circular inner orbit. (c) Sky separation of B and C at primary conjunctions around 2021, in units of $R_B$. A geometric eclipse occurs below the grazing limit, $1+R_C/R_B$. (d) The geometric factor of B's inner velocity semiamplitude, $\sin i_{\rm in}$ relative to its 2019--2023 mean, with the clock (dotted). All curves are drawn at the median and profile points.\label{fig:dyn}}
\end{figure*}

The inner relative orbit has a semimajor axis of 0.32\,mas (Figure~\ref{fig:orbit}a), and its line of nodes is rotated from the outer one by $\Omega_{\rm in}-\Omega_{\rm out}=23^\circ$ or $-29^\circ$. This rotation about the line of sight makes up almost all of the mutual inclination (Figure~\ref{fig:orbit}b).

The mass of A differs between the two branches by 1.8\ times the mean half-width of their 68\% intervals, and the osculating outer inclination at $T_{\rm ref}$ is $81.0^{+1.2}_{-1.2}$\,deg for $\Fp$ and $87.6^{+1.2}_{-1.2}$\,deg for $\Fm$. On both branches the orbits regress about the total angular momentum. Between the Gaia epochs and $T_{\rm ref}$, however, the outer inclination falls on $\Fp$ and rises on $\Fm$, so the same Gaia solution corresponds to different orbits in 2021.

\section{Why the Eclipses Switched Off} \label{sec:season}
\subsection{A two-year eclipse season} \label{sec:cone}
For a circular inner orbit, the disks of B and C overlap at conjunction only while the orbit lies within
\begin{equation}
|i_{\rm in}-90^\circ|<\arcsin\frac{R_B+R_C}{a_{\rm in}}=3.07^\circ
\label{eq:band}
\end{equation}
of edge-on, where $a_{\rm in}=18.7\,(R_B+R_C)$. The eclipsing band is thus 6.1$^\circ$ wide. At $T_{\rm ref}$ the inner orbit was 0.19$^\circ$ or 0.25$^\circ$ from edge-on.

Because the orbital planes are misaligned, A exerts a torque on the inner orbit, and both angular momenta precess about their sum, $\boldsymbol{J}=\boldsymbol{L}_{\rm in}+\boldsymbol{L}_{\rm out}$ (Figures~\ref{fig:orbit}b and \ref{fig:dyn}a). For a nearly circular inner orbit at quadrupole order the precession is uniform \citep{Soderhjelm1975,Torres2001,Zasche2012}. The inner inclination then follows a clock (Appendix~\ref{app:clock}),
\begin{equation}
\cos i_{\rm in}(t)=\cos\theta\cos I_J+\sin\theta\sin I_J\cos\frac{2\pi(t-t_0)}{P_{\rm node}},
\label{eq:clock}
\end{equation}
where $\theta$ is the angle between $\hat{L}_{\rm in}$ and $\boldsymbol{J}$, $I_J$ the angle between $\boldsymbol{J}$ and the direction to the observer, $t_0$ the time of minimum $i_{\rm in}$, and
\begin{equation}
P_{\rm node}\simeq\frac{4}{3}\,\frac{M_A+M_B+M_C}{M_A}\,\frac{P_{\rm out}^2}{P_{\rm in}}\,\frac{(1-e_{\rm out}^2)^{3/2}}{\cos i_{\rm mut}}\,\frac{L_{\rm out}}{J}.
\label{eq:pnode}
\end{equation}
The nodal period is the mean period at which $\hat{L}_{\rm in}$ circulates about $\boldsymbol{J}$, and successive seasons are separated alternately by a shorter and a longer gap. At the median points the three-body integrations give 48.7\ and 52.7\,yr, and Equation~(\ref{eq:pnode}) gives 48.6\ and 52.8\,yr. The integrated sampled states give $48.7^{+2.5}_{-2.3}$\,yr on $\Fp$ and $52.7^{+3.3}_{-2.8}$\,yr on $\Fm$. G3430's nodal period is about a tenth of those of SS~Lac and HS~Hya, whose eclipses also stopped \citep{Torres2001,Vokrouhlicky2022}. It is close to the 63.5\,yr of V907~Sco \citep{Zasche2023}, and the most compact triples precess within a few years \citep[e.g., KOI-126;][]{Yenawine2022}.

At the 2021 crossing $i_{\rm in}$ changes by 2.5$^\circ$\,yr$^{-1}$ or 2.9$^\circ$\,yr$^{-1}$, so the inner orbit crosses the band in about two years. A geometric eclipse is a primary conjunction with overlapping disks, detectable or not, and a season is a group of them less than 2\,yr apart. At the median points the 2021 season runs from 2020.7\ to 2023.1\ or from 2020.8\ to 2022.7\ (Figure~\ref{fig:dyn}b), and over the sampled states it lasts $2.4^{+0.3}_{-0.3}$ or $1.9^{+0.1}_{-0.1}$\,yr. By Sector~72 the orbit is 4.8$^\circ$ or 5.9$^\circ$ from edge-on, and B and C pass 2.6\ or 3.2\,$R_B$ apart at conjunction, beyond the grazing limit of 1.65\,$R_B$ (Figure~\ref{fig:dyn}c). The eclipses switched off because the inner orbit precessed out of the eclipsing band.

The 2021 eclipses themselves show this motion. We fitted them one by one, with the stars held at the median point. The closest approach of B and C grows from 0.05\ to 0.17\ of $R_B+R_C$ over the 68\,d (Figure~\ref{fig:ecl}e). With the stars held fixed, a constant geometry is disfavored at a nominal 4.6$\sigma$, beyond all 600\ simulated constant geometries (Appendix~\ref{app:impl}).

\subsection{What is modeled and what is observed} \label{sec:nbody}
All dates come from direct integrations of every primary conjunction from 1990 to 2080, and the clock reproduces them for all 12\ fitted points (Appendix~\ref{app:clock}).

TESS observed eclipses in late 2021 and none in late 2023. The model places the last geometric eclipse between them, in 2022.8--2023.1\ on $\Fp$ or 2022.6--2022.8\ on $\Fm$ (68\% intervals, Appendix~\ref{app:points}). Archival photometry samples few conjunctions (Figure~\ref{fig:ecl}f). One ASAS-SN bin, at 2021.08, lies within an hour of a primary conjunction inside the modeled season for every fitted point. It is $41\pm10$\,mmag fainter than the mean, where the median points predict 37\ and 35\,mmag, a tentative detection. A single ASAS-SN visit, on 2016 January 29, is 0.13\,mag faint when no fitted orbit eclipses. It has no entry in the survey's image-subtraction light curve \citep{Jayasinghe2018}, remains unconfirmed and does not influence the fit.

\subsection{Predictions} \label{sec:pred}
The shorter gap between seasons follows 2021 in $\Fp$ and the longer in $\Fm$, so geometric primary eclipses should resume in 2038--2041\ on $\Fp$ or 2047--2050\ on $\Fm$ (68\% intervals). With the night offsets of Section~\ref{sec:masses} they resume 1.5\ or 2.7\,yr later (Table~\ref{tab:sys}). B's inner velocity amplitude peaks at edge-on and then falls with $\sin i_{\rm in}$, by $2.6^{+0.8}_{-0.7}\%$ on $\Fp$ or $7.1^{+1.1}_{-1.0}\%$ on $\Fm$ by 2030 (Figure~\ref{fig:dyn}d), about 1.0\ or 2.9\,km\,s$^{-1}$ at the median points. The two predicted falls differ by 1.9\,km\,s$^{-1}$, so measuring the fall to about 0.5\,km\,s$^{-1}$ would separate the branches. This precision requirement is an estimate made without simulating the observations.

\section{Discussion and Conclusions} \label{sec:sum}
The architecture and the luminosity of the companion follow almost directly from the data. The inner node does not. It is constrained only when the eclipse times, the eclipse shapes and the switch-off are fitted with the same three-body motion, which is what the joint fit adds. The switch-off makes a slow secular motion measurable within two years. The hot Jupiter TOI-2535\,b is a planetary counterpart: its transits stopped as the transit chord drifted, and its perturber is inferred \citep{Yee2026}. In G3430 the perturber is star A, with a measured mass and orbit.

The architecture and the 2021 eclipse season are insensitive to the treatment of the 36 blended spectra, but the inferred masses and forecasts are not (Table~\ref{tab:sys}). Assigning A's velocity to B at these epochs, with an enlarged error, limits the fit to velocity separations below about 20\,km\,s$^{-1}$ there (Appendix~\ref{app:impl}) and lowers the inferred masses. Because B is not resolved in these spectra and the separation at which they blend has not been calibrated, these proxy velocities are excluded from the reference fit.

The data do not yet decide between the branches. At the profile points of median $i_{\rm mut}$, the highest posterior densities found differ by $\Delta\ln\mathcal{P}=10.2$ in favor of $\Fp$, but this is not a ratio of branch probabilities. The intervals of dates and forecasts are therefore conditional on the branch (Appendix~\ref{app:points}).

None of the 12\ fitted points shows Lidov--Kozai cycles in 1200\,yr integrations. In the tail probes of Appendix~\ref{app:lk}, large-amplitude cycles set in between 40.1$^\circ$ and 43.1$^\circ$ ($\Fp$) and between 42.0$^\circ$ and 45.0$^\circ$ ($\Fm$). These onsets lie above the 95th percentiles of $i_{\rm mut}$ on both branches, so the switch-off is driven by nodal precession.

\noindent\textit{Conclusions.}\label{sec:concl}
\begin{enumerate}
\item G3430 is the hierarchical triple A+(B+C). Its eclipsing pair, A's luminous companion on the 291\,d astrometric orbit, outweighs A but emits only about a fifth of the $G$-band light.
\item A single three-body model jointly fits the Gaia astrometric orbital solution, eclipse photometry, radial velocities, and SED, constraining all three stellar masses without imposing stellar-evolution tracks or a mass–radius relation. The masses of A and B depend on how the blended spectra and nightly offsets are treated, and so do the mutual inclinations and forecasts below, which assume independent velocity errors (Table~\ref{tab:sys}).
\item Only the dynamics constrain the inner node. The fit finds two orientations, $\Fp$ and $\Fm$, of undetermined relative probability, with mutual inclinations of 25$^\circ$ and 30$^\circ$.
\item The eclipses switched off because nodal precession, with a 49\ or 53\,yr period, carried the inner orbit through the eclipsing band in about two years. Their return in 2038--2041\ on $\Fp$ or 2047--2050\ on $\Fm$, or the fall of $K_B$ by 2030, would identify the orientation.
\end{enumerate}
Because Gaia selects triples by their outer orbits, it can find misaligned triples like G3430 between their eclipse seasons.

\begin{acknowledgments}
We thank Dong Lai, Thomas Baycroft, Jun Yan Lau, Xiumin Huang, and Bryant Randolph for helpful discussions. This work is supported by the National Key R\&D Program of China (grant Nos.\ 2024YFA1611801 and 2024YFC2207700), by the National Natural Science Foundation of China (NSFC) under grant No.\ 12473066, by the Shanghai Jiao Tong University 2030 Initiative, and by the China--Chile Joint Research Fund (CCJRF No.\ 2205). CCJRF is provided by the Chinese Academy of Sciences South America Center for Astronomy (CASSACA) and established by the National Astronomical Observatories, Chinese Academy of Sciences (NAOC) and the Chilean Astronomy Society (SOCHIAS) to support China--Chile collaborations in astronomy. This project is supported in part by Office of Science and Technology, Shanghai Municipal Government (grant Nos.\ 24DX1400100, ZJ2023-ZD-001). The computations in this paper were run on the Siyuan-1 and $\pi$ 2.0 clusters supported by the Center for High Performance Computing at Shanghai Jiao Tong University. This work made use of the data from LAMOST (Large Sky Area Multi-Object Fiber Spectroscopic Telescope, also known as the Guoshoujing Telescope) (\url{https://cstr.cn/31118.02.LAMOST}). LAMOST is a Chinese national mega-science facility, operated by National Astronomical Observatories, Chinese Academy of Sciences. This work has made use of data from the European Space Agency (ESA) mission Gaia (\url{https://www.cosmos.esa.int/gaia}), processed by the Gaia Data Processing and Analysis Consortium (DPAC, \url{https://www.cosmos.esa.int/web/gaia/dpac/consortium}); funding for the DPAC has been provided by national institutions, in particular the institutions participating in the Gaia Multilateral Agreement. This paper includes data collected by the TESS mission, funded by the NASA Science Mission Directorate and publicly available from the Mikulski Archive for Space Telescopes (MAST). The TESS-SPOC and QLP light curves used in this paper can be found in MAST: \dataset[doi:10.17909/t9-wpz1-8s54]{https://doi.org/10.17909/t9-wpz1-8s54} and \dataset[doi:10.17909/t9-r086-e880]{https://doi.org/10.17909/t9-r086-e880}. We use data from the ASAS-SN Sky Patrol, the WASP project (SuperWASP) and the KELT survey. This publication makes use of data products from the Two Micron All Sky Survey, which is a joint project of the University of Massachusetts and the Infrared Processing and Analysis Center/California Institute of Technology, funded by the National Aeronautics and Space Administration and the National Science Foundation. This research was made possible through the use of the AAVSO Photometric All-Sky Survey (APASS), funded by the Robert Martin Ayers Sciences Fund and NSF AST-1412587. This publication makes use of data products from the Wide-field Infrared Survey Explorer, a joint project of the University of California, Los Angeles, and the Jet Propulsion Laboratory/California Institute of Technology, funded by NASA.
\end{acknowledgments}
\facilities{LAMOST, Gaia, TESS, ASAS-SN, SuperWASP, KELT, WISE, CTIO:2MASS, FLWO:2MASS}
\software{astropy \citep{Astropy2013,Astropy2018,Astropy2022}, batman \citep{Kreidberg2015}, REBOUND \citep{Rein2012,Rein2015}, REBOUNDx \citep{Tamayo2020,Lu2023}, MIST \citep{Choi2016,Dotter2016}, NumPy \citep{Harris2020}, SciPy \citep{Virtanen2020}, Matplotlib \citep{Hunter2007}}

\restartappendixnumbering
\appendix
\twocolumngrid

\section{Implementation and Checks} \label{app:impl}
\textit{Likelihood.} The likelihood is the product of seven terms: the two velocity sets, the 2021 and 2023 TESS fluxes, the archival bins, the seven Gaia quantities and the ten magnitudes. Table~\ref{tab:full} lists the priors; that on [Fe/H] is from LAMOST. The inner elements are those of C about B, with argument of periastron $\omega_B+180^\circ$, and $T_{\rm c,in}$ is the two-body time of primary conjunction. The outer elements are those of A about the B+C barycenter. The integration uses IAS15 through the 2021 TESS windows and WHFast (step $P_{\rm in}/120$) elsewhere. At 8\ states per branch, halving the step or using IAS15 throughout changes $\ln\mathcal{L}$ by at most 0.032.

\textit{Velocities.} Each set has a Gaussian likelihood with a jitter and its own systemic velocity, marginalized with a flat prior. The errors are 1.5\,km\,s$^{-1}$ for A and 1.1--6.5\,km\,s$^{-1}$ for the 28\ resolved velocities of B. The 36\ blended exposures enter with A's velocity only, from the same two-component broadening-function fit, which we test within the fit rather than by injection into the spectra. A pull of A's velocity toward B by a fraction $\beta$ of the model separation is limited to $\beta\le0.08$ (95\%), whereas the light-weighted mean of a blend would give about 0.19. Allowing $0\le\beta\le0.3$ moves the medians by at most 0.5\,$h$, where $h$ is the half-width of the 68\% interval of the reference fit (as in Table~\ref{tab:sys}). The residuals of A and B are correlated within a night, which we model with a common offset for each night (local noon to noon, UTC+8), Gaussian with dispersion $\tau$ and marginalized analytically, with equal prior weights on $\tau=0$, 0.25, \ldots, 4\,km\,s$^{-1}$. Reweighting the 18{,}000 states of the fit by the ratio of the two velocity likelihoods gives approximate Bayes factors of $e^{4.9}$ or $e^{8.8}$ in favor of these offsets and median dispersions $\tau=1.25$ or 1.50\,km\,s$^{-1}$. The effective sample sizes of the weights are 6.7\% and 0.7\%, so the $\Fm$ values are only indicative. Sampled like the fit, the posterior with these offsets gives $i_{\rm mut}=26.1^{+3.2}_{-2.5}$ or ${33.5^{+5.2}_{-4.1}}^\circ$ (masses in Section~\ref{sec:masses}, other shifts in Table~\ref{tab:sys}); its forecast intervals are up to 1.7\ times wider. For the masses and $i_{\rm mut}$ its runs have $\hat{R}\le1.03$ and 1.14\ and differ by up to 16\% of $h$; over all quantities $\hat{R}$ reaches 1.06 and 1.33, the latter for the inner eccentricity. This incomplete mixing limits the precision of the summaries of this posterior. The intervals of the reference fit remain conditional on independent velocity errors. At the blended epochs the fit keeps B within $21.8^{+2.2}_{-2.0}$ or $18.1^{+1.5}_{-1.1}$\,km\,s$^{-1}$ of A, below the smallest resolved separation, 32.8\,km\,s$^{-1}$. The separation at which these spectra blend is not calibrated, however, and a limit of 20\,km\,s$^{-1}$ gives $M_A=1.15^{+0.08}_{-0.12}\,M_\odot$ on $\Fp$ (Section~\ref{sec:sum}).

\textit{TESS, 2021.} The flux at time $t$ in sector $s$ is
\begin{equation}
F(t)=b(t)\left\{1-d_s\left[f_B\lambda_B(t)+f_C\lambda_C(t)\right]\right\},
\label{eq:flux}
\end{equation}
where $\lambda_B$ and $\lambda_C$ are the blocked fractions of the light of B and C, $f_B$ and $f_C$ their TESS-band light shares, $d_s$ a depth scale (3\% prior), and $b(t)$ a quadratic baseline for each window, marginalized analytically with a flat prior. The blocked fractions, for quadratic limb darkening \citep{Kipping2013}, come from an occultation routine checked against batman \citep{Kreidberg2015} and are averaged over the 600\,s exposures. Within a window the inner orbit is advanced from the integrated state by a two-body expansion, at the emission time of the B+C barycenter. At the 8\ states above, three-body positions change the fluxes by at most 2.1\,ppm, against errors of about 1070\,ppm. The model omits the light-travel time across the inner orbit, which shifts the primary eclipses by 20--30\,s and the secondaries by 6--9\,s in the opposite sense. With every star at its own emission time, a refit of $P_{\rm in}$, $T_{\rm c,in}$ and $\sqrt{e_{\rm in}}\cos\omega_B$ at the median points recovers the $\ln\mathcal{L}$ of the same refit without the light time to within 0.2. It lowers $e_{\rm in}\cos\omega_B$ by 25\% of its half-width, delays $T_{\rm c,in}$ by 24--32\,s and shortens $P_{\rm in}$ by about 2\,s, so $T_{\rm c,in}$ is a coordinate of this timing convention.

\textit{Each eclipse alone.} Each of the eleven windows was fitted alone with the stars at the median point and three free quantities: the closest approach $x$ of B and C, the mid-time and the depth (Figure~\ref{fig:ecl}d, e). A linear trend of $x$ in time is preferred to a constant by $\Lambda=20.9$ (20.3\ with the $\Fm$ stars), a nominal 4.6$\sigma$ (4.5$\sigma$) for $\chi^2_1$. In 600\ simulations of the best constant geometry, with Gaussian noise or cyclically shifted residuals and refitted as the data, $\Lambda$ exceeds 6.63 in 1.0\% and 1.3\% of the cases, respectively, as expected for $\chi^2_1$, and reaches at most 12.7, below the observed value. For either noise model this bounds the false-alarm probability below 1\% (95\% confidence) but does not calibrate the nominal 4.6$\sigma$. With the stars of 24\ posterior states, $\Lambda=16.3$--26.3. The depths scatter by 1.9\% (primary) and 4.2\% (secondary) rms, against formal errors of 0.3\% and 1.4\%. In the event-to-event scatter test, the flux in window $j$ is modeled as $F_j(t)=b_j(t)E_j(t)+\delta_j[E_j(t)-1]$, where $E_j(t)$ is the diluted eclipse model before multiplication by the baseline. The $\delta_j$ are independent Gaussian variables with zero mean and dispersion $\sigma_{\rm ev}$, marginalized jointly and analytically with the baseline coefficients. The prior on $\sigma_{\rm ev}$ is half-normal with scale 0.05, truncated to $[0,0.2]$. Added to the night-offset fit, it gives $\sigma_{\rm ev}={2.5^{+1.0}_{-0.7}}\%$ of the depth on both branches. It moves the masses, $i_{\rm mut}$ and the forecasts by at most 0.3\ half-widths and the radii by up to 0.5\ (Table~\ref{tab:sys}). Its $\Fm$ runs agree least well, with $\hat{R}\le1.20$ for the masses and $i_{\rm mut}$ but up to 1.58\ for the inner eccentricity.

\textit{TESS, 2023, and archival photometry.} For each Sector~72 window the term is the log ratio of the likelihoods of the eclipse model and of a flat light curve, each with its own quadratic baseline. The denominator does not depend on the trial, so a trial without eclipses in 2023 is neither rewarded nor penalized. The search of the whole sector recovers injected eclipses of the 2021 shape 0.4\% deep or deeper in at least 99\% of 400\ trials, and 80\% of those 0.2\% deep, so the deepest dip found, 0.20\%, is not a completeness limit. For an archival bin with predicted dimming $m$, the term is the log ratio of a mixture likelihood at $m$ and at zero (0.95 on a Gaussian of the survey's scatter, 0.05 on one five times broader). SuperWASP is treated as $V$ and KELT as $R$.

\textit{Gaia.} The 30\ transits of 2014.7--2017.3, with their scan angles and parallax factors, are those of the Gaia Observation Forecast Tool. The along-scan photocenter coordinates of Equation~(\ref{eq:phot}) are fitted by least squares with five astrometric parameters and a Thiele--Innes orbit, whose period, eccentricity and periastron time are refined by Gauss--Newton iteration from the osculating outer orbit at J2016.0. The fitted parallax term, the leakage of the orbit, is added to the trial parallax. The model vector $(A,B,F,G,P,e,\varpi)$, in mas except for $P$ in days and the dimensionless $e$, is compared with the Gaia DR3 catalog solution (\texttt{nss\_two\_body\_orbit}) through a multivariate Gaussian likelihood. Its covariance $C_{ij}=\sigma_i\sigma_j\rho_{ij}$ is the corresponding submatrix of the catalog covariance, whose correlation matrix is rebuilt from \texttt{corr\_vec} (upper triangle in column-major order, parameters ordered $(\alpha,\delta,\varpi,\mu_{\alpha*},\mu_\delta, A,B,F,G,e,P,T_p)$). The term is thus a likelihood for the catalog solution, not a fit to epoch astrometry. The periastron epoch is held out because the velocities fix the outer phase. The model value lies 1.5$\sigma$ or 1.7$\sigma$ from the epoch implied by the other seven quantities, and Table~\ref{tab:sys} gives the effect of including it. The catalog does not list its transits. Over 72\ alternative projections, with transits removed or reweighted, the medians move by up to 0.4\ half-widths and the parallax by up to 1.1\ (Table~\ref{tab:sys}).

\textit{SED.} The ten magnitudes of Section~\ref{sec:data} are Vega magnitudes from APASS DR9, Gaia DR3, 2MASS and AllWISE. The 2MASS and AllWISE quality flags are clean, and no other source lies within $18''$. The MIST bolometric corrections are interpolated linearly in $T_{\rm eff}$, $\log g$, [Fe/H] and $A_V$ ($R_V=3.1$), with error floors of 0.01\,mag (Gaia) and 0.03\,mag (others). No integrated state eclipses between 2003.4\ and 2020.0, which holds the APASS, Gaia and WISE epochs. The 2MASS scan (1997--2001) could have met an eclipse of the previous $\Fm$ season with a probability below 3\%.

\section{Parameters and Sampling} \label{app:full}\label{app:diag}\label{app:points}
\begin{deluxetable*}{llcc@{\hspace{2.0em}}lcc}[t!]
\tabletypesize{\scriptsize}
\tabcolsep=4pt
\tablecaption{The joint fit for the two node branches: priors and sampled posteriors of 29 of the 36\ coordinates (left), and derived quantities, sampled forecasts and representative solutions (right)\label{tab:full}\label{tab:sampled}}
\tablehead{\colhead{Sampled coordinate} & \colhead{Prior} & \colhead{$F_+$} & \colhead{$F_-$} & \colhead{Quantity} & \colhead{$F_+$} & \colhead{$F_-$}}
\startdata
$P_{\rm out}$ (d) & $\mathcal{U}(280,\,300)$ & $291.01^{+0.27}_{-0.27}$ & $290.93^{+0.31}_{-0.30}$ & \multicolumn{3}{c}{\textit{Derived from the samples}} \\
\cline{5-7}
$\sqrt{e_{\rm out}}\cos\omega_A$ & $\mathcal{U}(-0.95,\,0.95)$ & $0.463^{+0.011}_{-0.011}$ & $0.499^{+0.012}_{-0.012}$ & $i_{\rm mut}$ at $T_{\rm ref}$ (deg) & $24.72^{+2.47}_{-2.09}$ & $29.54^{+3.46}_{-3.04}$ \\
$\sqrt{e_{\rm out}}\sin\omega_A$ & $\mathcal{U}(-0.95,\,0.95)$ & $0.204^{+0.018}_{-0.019}$ & $0.198^{+0.019}_{-0.019}$ & $e_{\rm in}$ & $0.0102^{+0.0019}_{-0.0015}$ & $0.0117^{+0.0025}_{-0.0020}$ \\
$T_{\rm p,out}-2457000$ (d) & $\mathcal{U}(1100,\,1500)$ & $1281.99^{+1.37}_{-1.36}$ & $1279.55^{+1.29}_{-1.27}$ & $i_{\rm out}$ (deg) & $81.0^{+1.2}_{-1.2}$ & $87.6^{+1.2}_{-1.2}$ \\
\cline{5-7}
$\cos i_{\rm out}$ & $\mathcal{U}(-1,\,1)$ & $0.157^{+0.021}_{-0.021}$ & $0.042^{+0.022}_{-0.021}$ & \multicolumn{3}{c}{\textit{Every tenth retained state}} \\
\cline{5-7}
$\Omega_{\rm out}$ (deg) & $\mathcal{U}(-360,\,720)$ & $115.92^{+0.83}_{-0.82}$ & $116.77^{+0.83}_{-0.83}$ & $e_{\rm out}$ & $0.256^{+0.010}_{-0.010}$ & $0.289^{+0.011}_{-0.011}$ \\
$P_{\rm in}$ (d) & $\mathcal{U}(12.3,\,12.4)$ & $12.3657^{+0.0021}_{-0.0021}$ & $12.3530^{+0.0025}_{-0.0025}$ & $(M_B+M_C)/M_A$ & $1.17^{+0.11}_{-0.10}$ & $1.36^{+0.12}_{-0.10}$ \\
$\sqrt{e_{\rm in}}\cos\omega_B$ & $\mathcal{U}(-0.5,\,0.5)$ & $-0.0788^{+0.0053}_{-0.0055}$ & $-0.0813^{+0.0063}_{-0.0066}$ & $G$-band light of A (\%)$^{m}$ & $78.6^{+0.7}_{-0.7}$ & $77.6^{+0.8}_{-0.8}$ \\
$\sqrt{e_{\rm in}}\sin\omega_B$ & $\mathcal{U}(-0.5,\,0.5)$ & $-0.0631^{+0.0236}_{-0.0183}$ & $-0.0715^{+0.0264}_{-0.0210}$ & $G$-band light of B (\%)$^{m}$ & $19.7^{+0.7}_{-0.7}$ & $20.7^{+0.8}_{-0.7}$ \\
$T_{\rm c,in}-2459000$ (d) & $\mathcal{U}(470,\,485)$ & $478.35553^{+0.00032}_{-0.00031}$ & $478.35539^{+0.00028}_{-0.00028}$ & $G$-band light of C (\%)$^{m}$ & $1.69^{+0.05}_{-0.05}$ & $1.75^{+0.05}_{-0.05}$ \\
$i_{\rm in}$ (deg) & $\mathcal{U}(80,\,100)$ & $90.19^{+0.05}_{-0.05}$ & $89.75^{+0.06}_{-0.06}$ & photocenter orbit / orbit of A$^{m}$ & $0.602^{+0.013}_{-0.014}$ & $0.610^{+0.013}_{-0.014}$ \\
$\Omega_{\rm in}$ (deg) & $\mathcal{U}(-360,\,720)$ & $138.95^{+2.69}_{-2.39}$ & $87.31^{+3.08}_{-3.49}$ & main-sequence offset (mag)$^{m}$ & $-0.78^{+0.68}_{-0.47}$ & $0.32^{+0.70}_{-0.61}$ \\
\cline{5-7}
$M_A$ ($M_\odot$) & $\mathcal{U}(0.4,\,3)$ & $1.429^{+0.271}_{-0.230}$ & $1.062^{+0.173}_{-0.152}$ & \multicolumn{3}{c}{\textit{Integrated for 2304 sampled states}} \\
\cline{5-7}
$M_B$ ($M_\odot$) & $\mathcal{U}(0.3,\,2.5)$ & $1.080^{+0.148}_{-0.131}$ & $0.873^{+0.104}_{-0.095}$ & first eclipse of the 2021 season (yr) & $2020.74^{+0.03}_{-0.30}$ & $2020.81^{+0.00}_{-0.03}$ \\
$M_C$ ($M_\odot$) & $\mathcal{U}(0.2,\,1.2)$ & $0.588^{+0.015}_{-0.015}$ & $0.567^{+0.016}_{-0.016}$ & last eclipse of the 2021 season (yr) & $2023.11^{+0.03}_{-0.30}$ & $2022.70^{+0.07}_{-0.07}$ \\
$\varpi$ (mas) & $\mathcal{U}(2,\,3.2)$ & $2.614^{+0.021}_{-0.021}$ & $2.648^{+0.021}_{-0.021}$ & season length (yr) & $2.4^{+0.3}_{-0.3}$ & $1.9^{+0.1}_{-0.1}$ \\
\cline{1-4}
$\ln[J_A/(\mathrm{km\,s}^{-1})]$ & $\mathcal{U}(-5,\,3)$ & $-1.55^{+1.32}_{-2.32}$ & $-0.50^{+0.59}_{-2.67}$ & nodal period (yr) & $48.7^{+2.5}_{-2.3}$ & $52.7^{+3.3}_{-2.8}$ \\
$\ln[J_B/(\mathrm{km\,s}^{-1})]$ & $\mathcal{U}(-5,\,3)$ & $-2.70^{+1.62}_{-1.57}$ & $-2.64^{+1.64}_{-1.60}$ & next eclipse season begins (yr) & $2039.2^{+1.8}_{-1.6}$ & $2048.6^{+1.7}_{-1.6}$ \\
$T_{\mathrm{eff},A}$ (K) & $\mathcal{U}(4500,\,8000)$ & $6203^{+284}_{-247}$ & $6122^{+267}_{-235}$ & fall of $K_B$ by 2030 (\%) & $2.6^{+0.8}_{-0.7}$ & $7.1^{+1.1}_{-1.0}$ \\
\cline{5-7}
$T_{\mathrm{eff},B}$ (K) & $\mathcal{U}(4000,\,7500)$ & $5344^{+171}_{-154}$ & $5473^{+170}_{-154}$ & \multicolumn{3}{c}{\textit{Median points: fitted values without intervals}} \\
\cline{5-7}
$T_{\mathrm{eff},C}$ (K) & $\mathcal{U}(3000,\,5500)$ & $3969^{+94}_{-87}$ & $4039^{+91}_{-86}$ & $e_{\rm out}$ & $0.256$ & $0.288$ \\
$R_A$ ($R_\odot$) & $\mathcal{U}(0.5,\,3)$ & $1.261^{+0.037}_{-0.037}$ & $1.264^{+0.036}_{-0.037}$ & $(M_B+M_C)/M_A$ & $1.17$ & $1.36$ \\
$R_B$ ($R_\odot$) & $\mathcal{U}(0.4,\,1.6)$ & $0.865^{+0.027}_{-0.026}$ & $0.827^{+0.024}_{-0.023}$ & $G$-band light of A, B, C (\%)$^{m}$ & $78.5$, $19.8$, $1.7$ & $77.5$, $20.7$, $1.8$ \\
$R_C$ ($R_\odot$) & $\mathcal{U}(0.3,\,1)$ & $0.562^{+0.017}_{-0.016}$ & $0.532^{+0.014}_{-0.014}$ & photocenter orbit / orbit of A$^{m}$ & $0.60$ & $0.61$ \\
$[\mathrm{Fe/H}]$ (dex) & $\mathcal{N}(-0.12,\,0.15)$ on $[-0.9,\,0.45]$ & $-0.08^{+0.14}_{-0.14}$ & $-0.10^{+0.13}_{-0.14}$ & $\ln\mathcal{L}$ & $7263.3$ & $7249.3$ \\
$A_V$ (mag) & $\mathcal{N}(0.3,\,0.18)$ on $[0,\,1.5]$ & $0.32^{+0.15}_{-0.15}$ & $0.30^{+0.15}_{-0.14}$ & eclipse season (yr) & $2020.7$--$2023.1$ & $2020.8$--$2022.7$ \\
$d_{43}$ & $\mathcal{N}(1,\,0.03)$ on $[0.8,\,1.2]$ & $0.979^{+0.016}_{-0.016}$ & $0.974^{+0.017}_{-0.017}$ & nodal period (yr) & $49$ & $53$ \\
$d_{44}$ & $\mathcal{N}(1,\,0.03)$ on $[0.8,\,1.2]$ & $1.003^{+0.017}_{-0.016}$ & $0.999^{+0.017}_{-0.017}$ & next eclipse season begins (yr) & $2039$ & $2049$ \\
$d_{45}$ & $\mathcal{N}(1,\,0.03)$ on $[0.8,\,1.2]$ & $1.025^{+0.018}_{-0.018}$ & $1.020^{+0.018}_{-0.018}$ & & & \\
\enddata
\tablecomments{Values with intervals are pooled medians and central 68\% intervals (Section~\ref{sec:sampling}). $\mathcal{U}$ and $\mathcal{N}$ are uniform and normal priors, truncated to the stated range. Elements are osculating Jacobi elements at $T_{\rm ref}$, and times are BJD$_{\rm TDB}$. Each run also requires $e_{\rm out}<0.9$, $e_{\rm in}<0.3$ and its own sign of $\sin(\Omega_{\rm in}-\Omega_{\rm out})$. $P_{\rm in}$ is the osculating period, which the perturbation by A makes differ from the mean period of the 2021 primary eclipses, 12.34513\,d (Figure~\ref{fig:ecl}d). $T_{\rm c,in}$ is the two-body conjunction time in the timing convention of the model, which omits the light-travel time across the inner orbit (Appendix~\ref{app:impl}). The masses rest on the resolved velocities of B (Section~\ref{sec:sum}; sensitivities in Table~\ref{tab:sys}). $J_A$ and $J_B$ are velocity jitters and $d_{43}$--$d_{45}$ the sector depth scales. The seven coordinates not listed are the TESS-band limb-darkening coordinates $q_1$ and $q_2$ of B and of C \citep{Kipping2013}, with priors $\mathcal{U}(0,\,1)$, and the sector error scales $\ln s_{43}$--$\ln s_{45}$, with priors $\mathcal{U}(-1,\,1.5)$. On the right, the second block uses every tenth saved time of the retained chains, 599{,}184\ states per branch. Its main-sequence offset is the mean over the ten SED bands, with equal weights, of the magnitude of three mean-sequence dwarfs of the sampled masses minus the observed magnitude, negative where the dwarfs are brighter (Appendix~\ref{app:ms}). The third block follows 2304\ weighted states of each branch with REBOUND from 1990 to 2080 (Appendix~\ref{app:points}). Its dates are those of geometric primary eclipses, one inner period apart, grouped into seasons as defined in Section~\ref{sec:cone}. The last block gives the median point of each branch, the representative solution of Section~\ref{sec:sampling}. Its eclipse dates and nodal periods come from one integration each, and the block above gives their sampled intervals. The relative probability of the branches is undetermined. $^{m}$Uses stellar calibrations: the light shares and the photocenter ratio depend on the MIST bolometric corrections, and the main-sequence offset on the mean dwarf sequence (Appendix~\ref{app:ms}).}
\end{deluxetable*}

\begin{deluxetable*}{llccccccccc}[t!]
\tabletypesize{\scriptsize}
\tabcolsep=3.6pt
\tablecaption{Precision and sensitivity of the main quantities\label{tab:sys}}
\tablehead{\colhead{} & \colhead{} & \colhead{} & \multicolumn{2}{c}{Precision (\% of $h$)} & \multicolumn{6}{c}{Shift of the median under an alternative (units of $h$)} \\ \cline{4-5}\cline{6-11}
\colhead{Quantity} & \colhead{} & \colhead{$h$} & \colhead{MC error} & \colhead{runs} & \colhead{Gaia proj.} & \colhead{Gaia $t_p$} & \colhead{depth scatter} & \colhead{A toward B} & \colhead{night offsets} & \colhead{proxies}}
\startdata
$M_A$ ($M_\odot$) & $F_+$ & 0.251 & 0.7 & 4.4 & 0.33 & $-$0.01 & $-$0.07 & $+$0.40 & $-$0.70 & $-$1.20 \\
 & $F_-$ & 0.163 & 0.9 & 1.5 & 0.19 & $+$0.10 & $-$0.13 & $+$0.15 & $-$1.33 & $-$0.41 \\
$M_B$ ($M_\odot$) & $F_+$ & 0.140 & 0.7 & 5.0 & 0.31 & $+$0.02 & $-$0.13 & $+$0.46 & $-$0.72 & $-$1.22 \\
 & $F_-$ & 0.099 & 0.9 & 1.9 & 0.17 & $+$0.14 & $-$0.21 & $+$0.16 & $-$1.44 & $-$0.41 \\
$M_C$ ($M_\odot$) & $F_+$ & 0.015 & 0.7 & 4.0 & 0.24 & $+$0.20 & $-$0.22 & $+$0.39 & $-$0.43 & $-$0.60 \\
 & $F_-$ & 0.016 & 0.8 & 2.4 & 0.04 & $+$0.28 & $-$0.26 & $+$0.11 & $-$1.07 & $-$0.18 \\
$i_{\rm mut}$ (deg) & $F_+$ & 2.28 & 0.9 & 3.9 & 0.23 & $+$0.20 & $-$0.08 & $-$0.13 & $+$0.60 & $+$0.90 \\
 & $F_-$ & 3.25 & 0.9 & 4.4 & 0.06 & $+$0.21 & $+$0.01 & $-$0.04 & $+$1.21 & $+$0.18 \\
nodal period (yr) & $F_+$ & 2.4 & 5.4 & 24 & 0.36 & $+$0.15 & $-$0.11 & $-$0.24 & $+$0.90 & $+$1.43 \\
 & $F_-$ & 3.1 & 4.2 & 11 & 0.14 & $+$0.10 & 0.00 & $-$0.11 & $+$1.70 & $+$0.33 \\
next season begins (yr) & $F_+$ & 1.7 & 4.0 & 16 & 0.22 & $+$0.18 & $-$0.02 & $-$0.06 & $+$0.89 & $+$1.29 \\
 & $F_-$ & 1.6 & 1.9 & 2.1 & 0.17 & $+$0.06 & $-$0.06 & $-$0.19 & $+$1.66 & $+$0.25 \\
fall of $K_B$ by 2030 (\%) & $F_+$ & 0.7 & 4.2 & 14 & 0.17 & $+$0.18 & $-$0.13 & $-$0.10 & $+$0.61 & $+$0.72 \\
 & $F_-$ & 1.1 & 3.0 & 16 & 0.06 & $+$0.24 & $-$0.10 & $+$0.02 & $+$0.55 & $-$0.01 \\
\enddata
\tablecomments{$h$ is the half-width of the 68\% interval of the reference fit. MC error is the largest Monte Carlo standard error of its 16th, 50th and 84th percentiles, and runs is the largest difference between the three runs in these percentiles (Appendix~\ref{app:diag}). For the forecasts each run has 768 integrated states. The other columns give the shift of the median, alternative minus fit, in units of $h$. These shifts are sensitivities, not significances, and the quoted intervals do not include them. The depth-scatter and night-offset columns compare posterior medians of fits whose $\Fm$ runs are less well mixed (Appendix~\ref{app:impl}), so they describe the central values, not the tails. Gaia proj.: the largest absolute shift over 72\ alternative projections of the photocenter onto the catalog solution. Gaia $t_p$: the catalog periastron epoch included through its Gaussian conditional on the seven fitted quantities. Depth scatter: an event-to-event scatter of the eclipse depths added to the night-offset fit, relative to that fit. A toward B: A\textquoteright s velocities at the blended epochs drawn toward B by a fraction $\beta$ of the separation, flat in $0\le\beta\le0.3$. Night offsets: one velocity offset per night, common to A and B, sampled like the fit. Proxies: B given A\textquoteright s velocity at the blended epochs (Section~\ref{sec:sum}). Appendix~\ref{app:impl} describes each alternative.}
\end{deluxetable*}

\begin{figure*}[t!]
\centering
\includegraphics[width=0.46\textwidth]{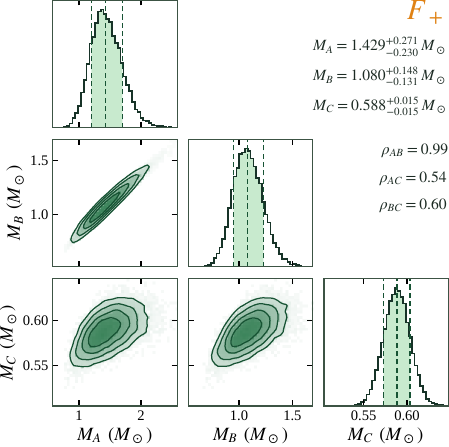}\hspace{0.04\textwidth}
\includegraphics[width=0.46\textwidth]{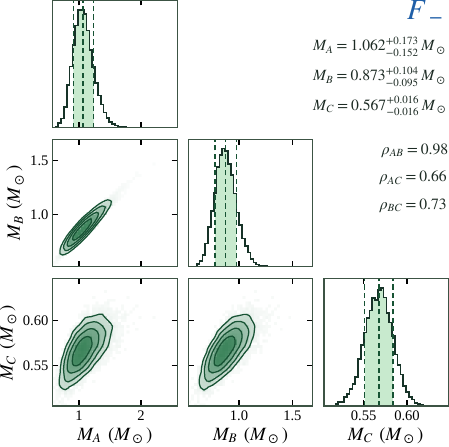}
\caption{Posterior of the three dynamical masses on $\Fp$ (left) and $\Fm$ (right), on common axes. Each run contributes 6,000 states, weighted so that the three runs count equally. Diagonal panels: histograms, with the 16th, 50th and 84th percentiles of all retained states (dashed). Other panels: density, with contours enclosing 39.3\%, 68.3\%, 86.5\% and 95.4\% of it. Upper right: the masses of Table~\ref{tab:full} and the correlation coefficients. Smooth contours are not a test of convergence (Table~\ref{tab:sys}).\label{fig:corner}}
\end{figure*}

The masses are correlated (Figure~\ref{fig:corner}): $M_A$ and $M_B$ both follow the outer velocity amplitude of the pair, and $M_A$ also correlates with $i_{\rm mut}$ ($-$0.52\ and $-$0.37).

\textit{Sampling.} Each run used 12\ temperatures up to $T=300$, 144\ walkers per temperature and 1500\ ladder-adaptation steps. Four in five proposals are differential-evolution moves \citep{terBraak2006}, the others stretch moves. Each run has its own seeds and a starting ensemble about points optimized at fixed node difference, wider in $i_{\rm mut}$ than the posterior and confined to one branch, so the runs do not exclude further modes. Steps 10{,}670--80{,}000\ are retained (1{,}996{,}848\ states per run). Because the walkers interact, the rank-normalized split $\hat{R}$ is computed over the three runs for the trajectory of each run's ensemble mean and of the fraction of walkers below a pooled percentile. The Monte Carlo error of a percentile is that of this fraction (overlapping batch means) divided by the posterior density, and $N/\tau$ is the retained length in autocorrelation times of the fraction. For the masses and $i_{\rm mut}$, $\hat{R}\le1.016$ and 1.036, and $N/\tau\ge34$ and 30. Over all 41\ monitored quantities, $\hat{R}$ reaches 1.064\ and 1.098, and $N/\tau$ falls to 21\ and 16. The runs reproduce the 68\% intervals to the precision listed in Table~\ref{tab:sys}, at most 0.6\,yr for the nodal period and 0.3\,yr for the next season, but not the far tails or the relative weight of the branches.

\textit{Fitted points and forecasts.} The five profile points of each branch are at the node differences that give the 5th, 16th, 50th, 84th and 95th percentiles of $i_{\rm mut}$. The forecasts integrate 2304\ states per branch from 1990 to 2080, 128 from each of six equal time strata of each run, weighted by the share of the retained states in their stratum. All of them eclipse in 2021. On $\Fp$ the 2021 season ends either in 2022.63--2022.90\ or in 2023.04--2023.17\ (28\% and 72\% of the weight), because the inclination advances in steps at the outer periastron passages.

\section{The Analytic Precession Clock} \label{app:clock}
\begin{figure*}[t!]
\centering\includegraphics[width=0.86\textwidth]{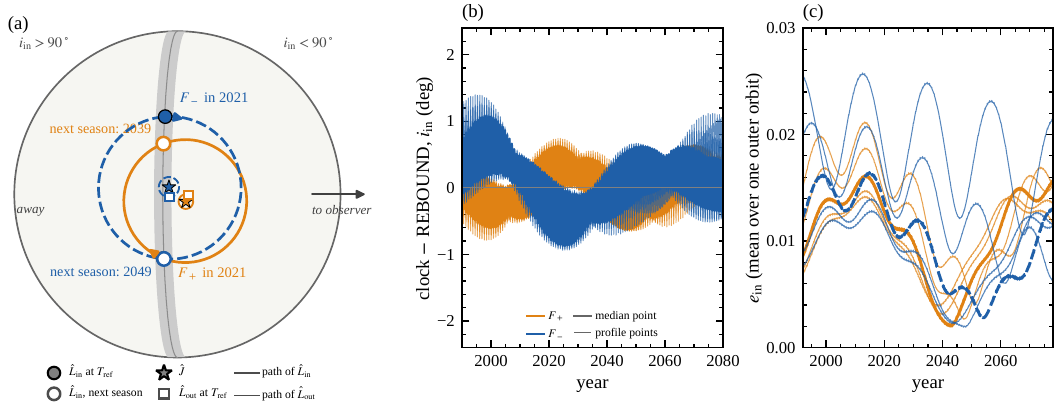}
\caption{The precession geometry to scale, and tests of the analytic clock. (a) Directions of the angular momenta of the two median points, seen along the mean $\hat{J}$. The inner pole $\hat{L}_{\rm in}$ moves on the large circles and $\hat{L}_{\rm out}$ on the small ones. The pair eclipses while $\hat{L}_{\rm in}$ lies in the gray band of Equation~(\ref{eq:band}). Arrows show the three years after $T_{\rm ref}$, and open circles the next crossing. (b) Clock minus REBOUND for $i_{\rm in}$ at the 12\ fitted points. The rms over 1990--2080 is at most 0.5$^\circ$, and the steps at outer periastron reach 1.4$^\circ$. (c) Inner eccentricity averaged over one outer orbit. All curves are drawn at the fitted points.\label{fig:clockgeom}}
\end{figure*}
\textit{Derivation.} For $L_{\rm in}\ll L_{\rm out}$ the inner pole precesses about $\hat{L}_{\rm out}$, and
\begin{equation}
P_{\rm node}\sim\frac{M_A+M_B+M_C}{M_A}\,\frac{P_{\rm out}^2}{P_{\rm in}},
\label{eq:scaling}
\end{equation}
40.6\ or 44.2\,yr at the median points. Here $L_{\rm in}/L_{\rm out}=0.12$ or 0.14, so both orbits precess about $\hat{J}$, and the quadrupole coefficient $4/3$ and the factors $(1-e_{\rm out}^2)^{3/2}$, $1/\cos i_{\rm mut}$ and $L_{\rm out}/J$ turn Equation~(\ref{eq:scaling}) into Equation~(\ref{eq:pnode}). The finite radii widen the edge-on condition to the band of Equation~(\ref{eq:band}), whose edges the fitted eccentricity moves by less than 1\%.

\textit{Both orbits and the gaps.} For a nearly circular inner orbit the double-averaged quadrupole interaction conserves $L_{\rm in}$, $L_{\rm out}$ and $i_{\rm mut}$. Both angular momenta therefore precess rigidly about $\boldsymbol{J}$, with regressing nodes \citep{Soderhjelm1975,Mazeh1976,Liu2015}, on cones of half-angles $\theta$ and $\alpha$ on opposite sides of $\boldsymbol{J}$ ($\theta+\alpha=i_{\rm mut}$). With $\phi$ the azimuth of $\hat{L}_{\rm in}$ about $\hat{J}$ from the projected line of sight, the inner orbit follows Equation~(\ref{eq:clock}) and the outer one
\begin{equation}
\cos i_{\rm out}=\cos\alpha\cos I_J-\sin\alpha\sin I_J\cos\phi.
\label{eq:cones}
\end{equation}
At the median points $\theta=22.2^\circ$ or $26.0^\circ$, $\alpha=2.6^\circ$ or $3.6^\circ$ and $I_J=81.9^\circ$ or $87.8^\circ$, so $i_{\rm out}$ oscillates by $\pm\alpha$ in antiphase with $i_{\rm in}$, causing its drift between the Gaia epochs and 2021. The inner orbit is edge-on at $\cos\phi_1=-\cot\theta\cot I_J$ and at $2\pi-\phi_1$, so successive band crossings are $(1-\phi_1/\pi)P_{\rm node}$ and $(\phi_1/\pi)P_{\rm node}$ apart. For $I_J<90^\circ$ the shorter gap follows a season in which $i_{\rm in}$ increases through $90^\circ$, as on $\Fp$; on $\Fm$ it decreases through $90^\circ$ in 2021 (Figure~\ref{fig:clockgeom}a). The crossings are 18.8\ and 29.9\,yr apart for $\Fp$ and 25.1\ and 27.7\,yr for $\Fm$.

\textit{Check against the integrations.} Started from the angular momenta at $T_{\rm ref}$, the clock reproduces the integrated nodal period to 0.6\%, the edges of the 2021 season within 0.3\,yr and the next season within 0.2\,yr at all 12\ points, whose orbit-averaged inner eccentricity stays below 0.026\ (Figure~\ref{fig:clockgeom}b, c). The clock fails where Lidov--Kozai cycles set in (Figure~\ref{fig:lk}b).

\section{Lidov--Kozai Test} \label{app:lk}
\begin{figure*}[t!]
\centering\includegraphics[width=0.86\textwidth]{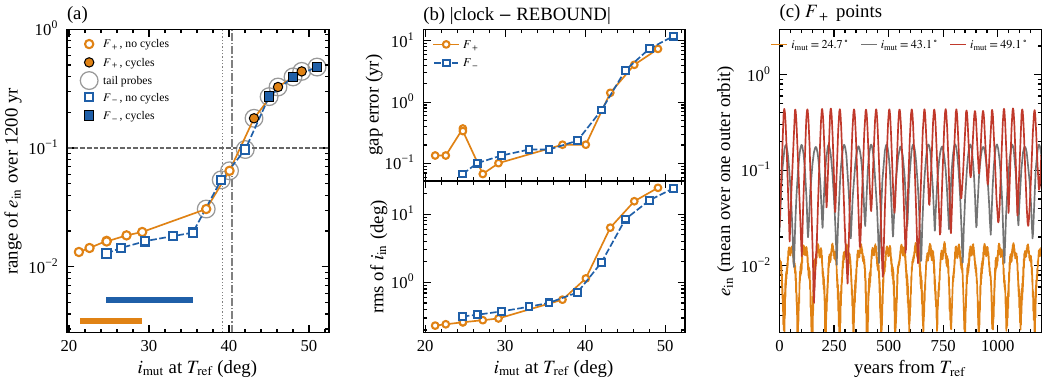}
\caption{Lidov--Kozai test with direct integrations of fitted points. (a) Range of the orbit-averaged inner eccentricity over 1200\,yr against $i_{\rm mut}$ at $T_{\rm ref}$. Rings mark the tail probes, and bars the sampled 5th to 95th percentiles of $i_{\rm mut}$. Lines mark a range of 0.1 (dashed), the test-particle threshold (dotted) and the threshold for a finite angular-momentum ratio (dash-dotted). (b) Error of the analytic clock for the gap to the next season and for $i_{\rm in}$. (c) Eccentricity histories of three $\Fp$ points. In the cycling points $e_{\rm in}$ and $i_{\rm mut}$ are anticorrelated (correlation $\le-0.96$).\label{fig:lk}}
\end{figure*}
We integrated the 12\ fitted points and ten tail probes for 1200\,yr to look for Lidov--Kozai cycles \citep{Kozai1962,Lidov1962,Naoz2016} (Figure~\ref{fig:lk}). The tail probes hold the node difference beyond the sampled 95th percentile and re-optimize the other coordinates. They locate the onset of cycles and are not probable configurations. A point cycles if its orbit-averaged inner eccentricity ranges over at least 0.1 in at least three cycles anticorrelated with $i_{\rm mut}$. This criterion marks large-amplitude cycles, not the linear instability.

No fitted point cycles (range at most 0.020). The onset brackets of Section~\ref{sec:sum} are consistent with the test-particle threshold of 39.2$^\circ$, with its value for a finite angular-momentum ratio, 40.3$^\circ$ \citep{Anderson2017}, and with a further rise from single-averaging terms \citep{Luo2016,Grishin2018}. Relativistic precession, tides and rotation, which we omit, can suppress the cycles \citep{Fabrycky2007,Naoz2016}. The sampled 95th percentiles of $i_{\rm mut}$, $29.15^\circ$ and $35.50^\circ$, lie below both brackets. Fewer than 0.2\% of the states lie above their lower edges, a fraction the runs do not establish precisely. Whether Lidov--Kozai cycles with tidal friction \citep{Fabrycky2007} or a primordial misalignment \citep{Moe2018} produced this configuration remains open.

\section{Stellar-model Check} \label{app:ms}
Each sampled state $k$ receives the weight
\begin{equation}
w_k=\sum_a p(a)\prod_{i}N\!\left(\ln\frac{R_{i,k}}{R_{{\rm iso},i}};s_R\right)N\!\left(\ln\frac{T_{i,k}}{T_{{\rm iso},i}};s_T\right),
\label{eq:mist}
\end{equation}
where $i$ runs over A, B and C, and $R_{{\rm iso},i}$ and $T_{{\rm iso},i}$ are the radius and temperature of a MIST isochrone of age $a$ at the mass of star $i$ and the [Fe/H] of the state, taken as initial (version 2.5, $v/v_{\rm crit}=0.4$, solar-scaled; \citealt{Dotter2026}). The function $N(x;s)$ is a normal density of zero mean, and $p(a)$ is flat from 0.3\ to 12.6\,Gyr, with one age for the three stars. The tolerances $s_R=4\%$ and $s_T=2.5\%$, for model accuracy and the inflation of B and C by tides and activity, are assumptions.

The ratio of the posterior mean weights, 3.5\ in favor of $\Fm$, is the factor by which an isochrone prior of weight $w$ would change the unknown ratio of the branch probabilities, because the stellar coordinates have the same priors on both branches, and we call it a consistency factor. It is 3.2--3.9\ in the three runs, and 4.1\ and 3.2\ for half and twice the tolerances. The effective numbers of weighted states are 6\% and 20\% of the 18{,}000.

Conditional on the data and the isochrones together, the weighted states give the masses of Section~\ref{sec:masses} at $3.2^{+2.5}_{-1.7}$\,Gyr on $\Fp$ and $5.0^{+2.1}_{-2.5}$\,Gyr on $\Fm$, with $M_A=$ 1.13, 1.15, 1.15\,$M_\odot$ in the three $\Fp$ runs. The main-sequence offset of Table~\ref{tab:full} uses the mean dwarf sequence of \citet{Pecaut2013}. The check assumes three coeval single stars, so we use it neither to weight the branches nor to replace the dynamical masses.

\bibliography{letter_3430}{}

@article{Rucinski2002,
  author = {Rucinski, Slavek M.},
  title = {Radial Velocity Studies of Close Binary Stars. VII. Methods and Uncertainties},
  journal = {\aj},
  year = {2002},
  volume = {124},
  number = {3},
  pages = {1746--1756},
  doi = {10.1086/342342},
  url = {https://doi.org/10.1086/342342},
}

@article{Halbwachs2023,
  author = {{Halbwachs}, Jean-Louis and {Pourbaix}, Dimitri and {Arenou}, Fr{\'e}d{\'e}ric and {Galluccio}, Laurent and {Guillout}, Patrick and {Bauchet}, Nathalie and {Marchal}, Olivier and {Sadowski}, Gilles and {Teyssier}, David},
  title = {{Gaia Data Release 3. Astrometric binary star processing}},
  journal = {\aap},
  year = {2023},
  volume = {674},
  pages = {A9},
  doi = {10.1051/0004-6361/202243969},
  url = {https://doi.org/10.1051/0004-6361/202243969},
  adsurl = {https://ui.adsabs.harvard.edu/abs/2023A&A...674A...9H},
  archiveprefix = {arXiv},
  eprint = {2206.05726},
  primaryclass = {astro-ph.SR},
}

@article{GaiaArenou2023,
  author = {{Gaia Collaboration} and {Arenou}, F. and {Babusiaux}, C. and {Barstow}, M.~A. and {Faigler}, S. and {Jorissen}, A. and {Kervella}, P. and {Mazeh}, T. and {Mowlavi}, N. and {Panuzzo}, P. and {Sahlmann}, J. and {Shahaf}, S. and {Sozzetti}, A. and {Bauchet}, N. and {Damerdji}, Y. and {Gavras}, P. and {Giacobbe}, P. and {Gosset}, E. and {Halbwachs}, J.-L. and {Holl}, B. and {Lattanzi}, M.~G. and {Leclerc}, N. and {Morel}, T. and {Pourbaix}, D. and {Re Fiorentin}, P. and {Sadowski}, G. and {S{\'e}gransan}, D. and {Siopis}, C. and {Teyssier}, D. and {Zwitter}, T. and {Planquart}, L. and {Brown}, A.~G.~A. and {Vallenari}, A. and {Prusti}, T. and {de Bruijne}, J.~H.~J. and {Biermann}, M. and {Creevey}, O.~L. and {Ducourant}, C. and {Evans}, D.~W. and {Eyer}, L. and {Guerra}, R. and {Hutton}, A. and {Jordi}, C. and {Klioner}, S.~A. and {Lammers}, U.~L. and {Lindegren}, L. and {Luri}, X. and {Mignard}, F. and {Panem}, C. and {Randich}, S. and {Sartoretti}, P. and {Soubiran}, C. and {Tanga}, P. and {Walton}, N.~A. and {Bailer-Jones}, C.~A.~L. and {Bastian}, U. and {Drimmel}, R. and {Jansen}, F. and {Katz}, D. and {van Leeuwen}, F. and {Bakker}, J. and {Cacciari}, C. and {Casta{\~n}eda}, J. and {De Angeli}, F. and {Fabricius}, C. and {Fouesneau}, M. and {Fr{\'e}mat}, Y. and {Galluccio}, L. and {Guerrier}, A. and {Heiter}, U. and {Masana}, E. and {Messineo}, R. and {Nicolas}, C. and {Nienartowicz}, K. and {Pailler}, F. and {Riclet}, F. and {Roux}, W. and {Seabroke}, G.~M. and {Sordo}, R. and {Th{\'e}venin}, F. and {Gracia-Abril}, G. and {Portell}, J. and {Altmann}, M. and {Andrae}, R. and {Audard}, M. and {Bellas-Velidis}, I. and {Benson}, K. and {Berthier}, J. and {Blomme}, R. and {Burgess}, P.~W. and {Busonero}, D. and {Busso}, G. and {C{\'a}novas}, H. and {Carry}, B. and {Cellino}, A. and {Cheek}, N. and {Clementini}, G. and {Davidson}, M. and {de Teodoro}, P. and {Nu{\~n}ez Campos}, M. and {Delchambre}, L. and {Dell'Oro}, A. and {Esquej}, P. and {Fern{\'a}ndez-Hern{\'a}ndez}, J. and {Fraile}, E. and {Garabato}, D. and {Garc{\'\i}a-Lario}, P. and {Haigron}, R. and {Hambly}, N.~C. and {Harrison}, D.~L. and {Hern{\'a}ndez}, J. and {Hestroffer}, D. and {Hodgkin}, S.~T. and {Jan{\ss}en}, K. and {Jevardat de Fombelle}, G. and {Jordan}, S. and {Krone-Martins}, A. and {Lanzafame}, A.~C. and {L{\"o}ffler}, W. and {Marchal}, O. and {Marrese}, P.~M. and {Moitinho}, A. and {Muinonen}, K. and {Osborne}, P. and {Pancino}, E. and {Pauwels}, T. and {Recio-Blanco}, A. and {Reyl{\'e}}, C. and {Riello}, M. and {Rimoldini}, L. and {Roegiers}, T. and {Rybizki}, J. and {Sarro}, L.~M. and {Smith}, M. and {Utrilla}, E. and {van Leeuwen}, M. and {Abbas}, U. and {{\'A}brah{\'a}m}, P. and {Abreu Aramburu}, A. and {Aerts}, C. and {Aguado}, J.~J. and {Ajaj}, M. and {Aldea-Montero}, F. and {Altavilla}, G. and {{\'A}lvarez}, M.~A. and {Alves}, J. and {Anders}, F. and {Anderson}, R.~I. and {Anglada Varela}, E. and {Antoja}, T. and {Baines}, D. and {Baker}, S.~G. and {Balaguer-N{\'u}{\~n}ez}, L. and {Balbinot}, E. and {Balog}, Z. and {Barache}, C. and {Barbato}, D. and {Barros}, M. and {Bartolom{\'e}}, S. and {Bassilana}, J.-L. and {Becciani}, U. and {Bellazzini}, M. and {Berihuete}, A. and {Bernet}, M. and {Bertone}, S. and {Bianchi}, L. and {Binnenfeld}, A. and {Blanco-Cuaresma}, S. and {Blazere}, A. and {Boch}, T. and {Bombrun}, A. and {Bossini}, D. and {Bouquillon}, S. and {Bragaglia}, A. and {Bramante}, L. and {Breedt}, E. and {Bressan}, A. and {Brouillet}, N. and {Brugaletta}, E. and {Bucciarelli}, B. and {Burlacu}, A. and {Butkevich}, A.~G. and {Buzzi}, R. and {Caffau}, E. and {Cancelliere}, R. and {Cantat-Gaudin}, T. and {Carballo}, R. and {Carlucci}, T. and {Carnerero}, M.~I. and {Carrasco}, J.~M. and {Casamiquela}, L. and {Castellani}, M. and {Castro-Ginard}, A. and {Chaoul}, L. and {Charlot}, P. and {Chemin}, L. and {Chiaramida}, V. and {Chiavassa}, A. and {Chornay}, N. and {Comoretto}, G.},
  title = {{Gaia Data Release 3. Stellar multiplicity, a teaser for the hidden treasure}},
  journal = {\aap},
  year = {2023},
  volume = {674},
  pages = {A34},
  doi = {10.1051/0004-6361/202243782},
  url = {https://doi.org/10.1051/0004-6361/202243782},
  adsurl = {https://ui.adsabs.harvard.edu/abs/2023A&A...674A..34G},
  archiveprefix = {arXiv},
  eprint = {2206.05595},
  primaryclass = {astro-ph.SR},
}

@article{GaiaVallenari2023,
  author = {{Gaia Collaboration} and {Vallenari}, A. and {Brown}, A.~G.~A. and {Prusti}, T. and {de Bruijne}, J.~H.~J. and {Arenou}, F. and {Babusiaux}, C. and {Biermann}, M. and {Creevey}, O.~L. and {Ducourant}, C. and {Evans}, D.~W. and {Eyer}, L. and {Guerra}, R. and {Hutton}, A. and {Jordi}, C. and {Klioner}, S.~A. and {Lammers}, U.~L. and {Lindegren}, L. and {Luri}, X. and {Mignard}, F. and {Panem}, C. and {Pourbaix}, D. and {Randich}, S. and {Sartoretti}, P. and {Soubiran}, C. and {Tanga}, P. and {Walton}, N.~A. and {Bailer-Jones}, C.~A.~L. and {Bastian}, U. and {Drimmel}, R. and {Jansen}, F. and {Katz}, D. and {Lattanzi}, M.~G. and {van Leeuwen}, F. and {Bakker}, J. and {Cacciari}, C. and {Casta{\~n}eda}, J. and {De Angeli}, F. and {Fabricius}, C. and {Fouesneau}, M. and {Fr{\'e}mat}, Y. and {Galluccio}, L. and {Guerrier}, A. and {Heiter}, U. and {Masana}, E. and {Messineo}, R. and {Mowlavi}, N. and {Nicolas}, C. and {Nienartowicz}, K. and {Pailler}, F. and {Panuzzo}, P. and {Riclet}, F. and {Roux}, W. and {Seabroke}, G.~M. and {Sordo}, R. and {Th{\'e}venin}, F. and {Gracia-Abril}, G. and {Portell}, J. and {Teyssier}, D. and {Altmann}, M. and {Andrae}, R. and {Audard}, M. and {Bellas-Velidis}, I. and {Benson}, K. and {Berthier}, J. and {Blomme}, R. and {Burgess}, P.~W. and {Busonero}, D. and {Busso}, G. and {C{\'a}novas}, H. and {Carry}, B. and {Cellino}, A. and {Cheek}, N. and {Clementini}, G. and {Damerdji}, Y. and {Davidson}, M. and {de Teodoro}, P. and {Nu{\~n}ez Campos}, M. and {Delchambre}, L. and {Dell'Oro}, A. and {Esquej}, P. and {Fern{\'a}ndez-Hern{\'a}ndez}, J. and {Fraile}, E. and {Garabato}, D. and {Garc{\'\i}a-Lario}, P. and {Gosset}, E. and {Haigron}, R. and {Halbwachs}, J.-L. and {Hambly}, N.~C. and {Harrison}, D.~L. and {Hern{\'a}ndez}, J. and {Hestroffer}, D. and {Hodgkin}, S.~T. and {Holl}, B. and {Jan{\ss}en}, K. and {Jevardat de Fombelle}, G. and {Jordan}, S. and {Krone-Martins}, A. and {Lanzafame}, A.~C. and {L{\"o}ffler}, W. and {Marchal}, O. and {Marrese}, P.~M. and {Moitinho}, A. and {Muinonen}, K. and {Osborne}, P. and {Pancino}, E. and {Pauwels}, T. and {Recio-Blanco}, A. and {Reyl{\'e}}, C. and {Riello}, M. and {Rimoldini}, L. and {Roegiers}, T. and {Rybizki}, J. and {Sarro}, L.~M. and {Siopis}, C. and {Smith}, M. and {Sozzetti}, A. and {Utrilla}, E. and {van Leeuwen}, M. and {Abbas}, U. and {{\'A}brah{\'a}m}, P. and {Abreu Aramburu}, A. and {Aerts}, C. and {Aguado}, J.~J. and {Ajaj}, M. and {Aldea-Montero}, F. and {Altavilla}, G. and {{\'A}lvarez}, M.~A. and {Alves}, J. and {Anders}, F. and {Anderson}, R.~I. and {Anglada Varela}, E. and {Antoja}, T. and {Baines}, D. and {Baker}, S.~G. and {Balaguer-N{\'u}{\~n}ez}, L. and {Balbinot}, E. and {Balog}, Z. and {Barache}, C. and {Barbato}, D. and {Barros}, M. and {Barstow}, M.~A. and {Bartolom{\'e}}, S. and {Bassilana}, J.-L. and {Bauchet}, N. and {Becciani}, U. and {Bellazzini}, M. and {Berihuete}, A. and {Bernet}, M. and {Bertone}, S. and {Bianchi}, L. and {Binnenfeld}, A. and {Blanco-Cuaresma}, S. and {Blazere}, A. and {Boch}, T. and {Bombrun}, A. and {Bossini}, D. and {Bouquillon}, S. and {Bragaglia}, A. and {Bramante}, L. and {Breedt}, E. and {Bressan}, A. and {Brouillet}, N. and {Brugaletta}, E. and {Bucciarelli}, B. and {Burlacu}, A. and {Butkevich}, A.~G. and {Buzzi}, R. and {Caffau}, E. and {Cancelliere}, R. and {Cantat-Gaudin}, T. and {Carballo}, R. and {Carlucci}, T. and {Carnerero}, M.~I. and {Carrasco}, J.~M. and {Casamiquela}, L. and {Castellani}, M. and {Castro-Ginard}, A. and {Chaoul}, L. and {Charlot}, P. and {Chemin}, L. and {Chiaramida}, V. and {Chiavassa}, A. and {Chornay}, N. and {Comoretto}, G. and {Contursi}, G. and {Cooper}, W.~J. and {Cornez}, T. and {Cowell}, S. and {Crifo}, F. and {Cropper}, M. and {Crosta}, M. and {Crowley}, C. and {Dafonte}, C. and {Dapergolas}, A. and {David}, M. and {David}, P. and {de Laverny}, P. and {De Luise}, F. and {De March}, R.},
  title = {{Gaia Data Release 3. Summary of the content and survey properties}},
  journal = {\aap},
  year = {2023},
  volume = {674},
  pages = {A1},
  doi = {10.1051/0004-6361/202243940},
  url = {https://doi.org/10.1051/0004-6361/202243940},
  adsurl = {https://ui.adsabs.harvard.edu/abs/2023A&A...674A...1G},
  archiveprefix = {arXiv},
  eprint = {2208.00211},
  primaryclass = {astro-ph.GA},
}

@article{Guo2025,
  author = {Guo, Sufen and Kovalev, Mikhail and Li, Jiao and Lü, Guoliang and Jia, Shi and Li, Zhenwei and Li, Jiangdan and Xiong, Jianping and Yang, Mingkuan and He, Tongyu and Chen, Xuefei and Han, Zhanwen},
  title = {Orbital Parameters of 665 Double-lined Spectroscopic Binaries in the LAMOST Medium-resolution Survey},
  journal = {\apjs},
  year = {2025},
  volume = {278},
  number = {2},
  pages = {46},
  doi = {10.3847/1538-4365/adced1},
  url = {https://doi.org/10.3847/1538-4365/adced1},
}

@article{Borkovits2016,
  author = {Borkovits, T. and Hajdu, T. and Sztakovics, J. and Rappaport, S. and Levine, A. and Bíró, I. B. and Klagyivik, P.},
  title = {A comprehensive study of the Kepler triples via eclipse timing},
  journal = {\mnras},
  year = {2016},
  volume = {455},
  number = {4},
  pages = {4136--4165},
  doi = {10.1093/mnras/stv2530},
  url = {https://doi.org/10.1093/mnras/stv2530},
  eprint = {https://academic.oup.com/mnras/article-pdf/455/4/4136/13772144/stv2530.pdf},
}

@article{Borkovits2025,
  author = {Borkovits, T. and Rappaport, S.~A. and Mitnyan, T. and B{\'\i}r{\'o}, I.~B. and Cs{\'a}nyi, I. and Forg{\'a}cs-Dajka, E. and Forr{\'o}, A. and Hajdu, T. and Seli, B. and Sztakovics, J. and G{\"o}bly{\"o}s, A. and P{\'a}l, A.},
  title = {{Then and now: A new look at the eclipse timing variations of hierarchical triple star candidates in the primordial Kepler field, revisited by TESS}},
  journal = {\aap},
  year = {2025},
  volume = {695},
  pages = {A209},
  doi = {10.1051/0004-6361/202453616},
  url = {https://doi.org/10.1051/0004-6361/202453616},
  adsurl = {https://ui.adsabs.harvard.edu/abs/2025A&A...695A.209B},
  archiveprefix = {arXiv},
  eprint = {2502.09480},
  primaryclass = {astro-ph.SR},
}

@article{Borkovits2026,
  author = {Borkovits, Tamás and Mitnyan, Tibor and Czavalinga, Donát R. and Rappaport, Saul A.},
  title = {A Few Years Later: Revisiting Period Variations of Eclipsing Binaries in the Northern Continuous Viewing Zone of TESS},
  journal = {Universe},
  year = {2026},
  volume = {12},
  number = {4},
  pages = {107},
  doi = {10.3390/universe12040107},
  url = {https://doi.org/10.3390/universe12040107},
}

@article{Rein2012,
  author = {{Rein}, H. and {Liu}, S.-F.},
  title = {{REBOUND: an open-source multi-purpose N-body code for collisional dynamics}},
  journal = {\aap},
  year = {2012},
  volume = {537},
  pages = {A128},
  doi = {10.1051/0004-6361/201118085},
  url = {https://doi.org/10.1051/0004-6361/201118085},
  adsurl = {https://ui.adsabs.harvard.edu/abs/2012A&A...537A.128R},
  archiveprefix = {arXiv},
  eprint = {1110.4876},
  primaryclass = {astro-ph.EP},
}

@article{Rein2015,
  author = {Rein, Hanno and Spiegel, David S.},
  title = {ias15: a fast, adaptive, high-order integrator for gravitational dynamics, accurate to machine precision over a billion orbits},
  journal = {\mnras},
  year = {2015},
  volume = {446},
  number = {2},
  pages = {1424--1437},
  doi = {10.1093/mnras/stu2164},
  url = {https://doi.org/10.1093/mnras/stu2164},
  eprint = {https://academic.oup.com/mnras/article-pdf/446/2/1424/9381126/stu2164.pdf},
}

@article{Kreidberg2015,
  author = {Kreidberg, Laura},
  title = {batman: BAsic Transit Model cAlculatioN in Python},
  journal = {\pasp},
  year = {2015},
  volume = {127},
  number = {957},
  pages = {1161--1165},
  doi = {10.1086/683602},
  url = {https://doi.org/10.1086/683602},
}

@article{Ricker2015,
  author = {{Ricker}, George R. and {Winn}, Joshua N. and {Vanderspek}, Roland and {Latham}, David W. and {Bakos}, G{\'a}sp{\'a}r {\'A}. and {Bean}, Jacob L. and {Berta-Thompson}, Zachory K. and {Brown}, Timothy M. and {Buchhave}, Lars and {Butler}, Nathaniel R. and others},
  title = {{Transiting Exoplanet Survey Satellite (TESS)}},
  journal = {JATIS},
  year = {2015},
  volume = {1},
  pages = {014003},
  doi = {10.1117/1.JATIS.1.1.014003},
  archivePrefix = {arXiv},
  eprint = {1406.0151},
  primaryClass = {astro-ph.EP},
  adsurl = {https://scixplorer.org/abs/2015JATIS...1a4003R},
}

@article{Caldwell2020,
  author = {Caldwell, Douglas A. and Tenenbaum, Peter and Twicken, Joseph D. and Jenkins, Jon M. and Ting, Eric and Smith, Jeffrey C. and Hedges, Christina and Fausnaugh, Michael M. and Rose, Mark and Burke, Christopher},
  title = {TESS Science Processing Operations Center FFI Target List Products},
  journal = {RNAAS},
  year = {2020},
  volume = {4},
  number = {11},
  pages = {201},
  doi = {10.3847/2515-5172/abc9b3},
  url = {https://doi.org/10.3847/2515-5172/abc9b3},
}

@article{Huang2020,
  author = {Huang, Chelsea X. and Vanderburg, Andrew and Pál, Andras and Sha, Lizhou and Yu, Liang and Fong, Willie and Fausnaugh, Michael and Shporer, Avi and Guerrero, Natalia and Vanderspek, Roland and Ricker, George},
  title = {Photometry of 10 Million Stars from the First Two Years of TESS Full Frame Images: Part I},
  journal = {RNAAS},
  year = {2020},
  volume = {4},
  number = {11},
  pages = {204},
  doi = {10.3847/2515-5172/abca2e},
  url = {https://doi.org/10.3847/2515-5172/abca2e},
}

@article{Cui2012,
  author = {{Cui}, Xiang-Qun and {Zhao}, Yong-Heng and {Chu}, Yao-Quan and {Li}, Guo-Ping and {Li}, Qi and {Zhang}, Li-Ping and {Su}, Hong-Jun and {Yao}, Zheng-Qiu and {Wang}, Ya-Nan and {Xing}, Xiao-Zheng and others},
  title = {{The Large Sky Area Multi-Object Fiber Spectroscopic Telescope (LAMOST)}},
  journal = {RAA},
  year = {2012},
  volume = {12},
  number = {9},
  pages = {1197--1242},
  doi = {10.1088/1674-4527/12/9/003},
  adsurl = {https://scixplorer.org/abs/2012RAA....12.1197C},
}

@article{Torres2000,
  author = {Torres, Guillermo and Stefanik, Robert P.},
  title = {The Cessation of Eclipses in SS Lacertae: The Mystery Solved*},
  journal = {\aj},
  year = {2000},
  volume = {119},
  number = {4},
  pages = {1914--1929},
  doi = {10.1086/301290},
  url = {https://doi.org/10.1086/301290},
}

@article{Lacy1999,
  author = {Sandberg Lacy, Claud H. and Helt, Bodil E. and Vaz, Luiz Paulo R.},
  title = {V907 Scorpii: A Remarkable Binary Star Whose Eclipses Turn On and Off and On and Off*},
  journal = {\aj},
  year = {1999},
  volume = {117},
  number = {1},
  pages = {541--547},
  doi = {10.1086/300710},
  url = {https://doi.org/10.1086/300710},
}

@article{Zasche2012,
  author = {{Zasche}, P. and {Paschke}, A.},
  title = {{HS Hydrae about to turn off its eclipses}},
  journal = {\aap},
  year = {2012},
  volume = {542},
  pages = {L23},
  doi = {10.1051/0004-6361/201219392},
  url = {https://doi.org/10.1051/0004-6361/201219392},
  adsurl = {https://ui.adsabs.harvard.edu/abs/2012A&A...542L..23Z},
  archiveprefix = {arXiv},
  eprint = {1205.6321},
  primaryclass = {astro-ph.SR},
}

@article{Naoz2016,
  author = {Naoz, Smadar},
  title = {The Eccentric Kozai-Lidov Effect and Its Applications},
  journal = {\araa},
  year = {2016},
  volume = {54},
  pages = {441--489},
  doi = {10.1146/annurev-astro-081915-023315},
  url = {https://doi.org/10.1146/annurev-astro-081915-023315},
}

@article{Kozai1962,
  author = {{Kozai}, Yoshihide},
  title = {{Secular perturbations of asteroids with high inclination and eccentricity}},
  journal = {\aj},
  year = {1962},
  volume = {67},
  pages = {591--598},
  doi = {10.1086/108790},
  url = {https://doi.org/10.1086/108790},
  adsurl = {https://ui.adsabs.harvard.edu/abs/1962AJ.....67..591K},
}

@article{Lidov1962,
  author = {Lidov, M. L.},
  title = {The evolution of orbits of artificial satellites of planets under the action of gravitational perturbations of external bodies},
  journal = {P\&SS},
  year = {1962},
  volume = {9},
  number = {10},
  pages = {719--759},
  doi = {10.1016/0032-0633(62)90129-0},
  url = {https://doi.org/10.1016/0032-0633(62)90129-0},
}

@article{Xu2026,
  author = {Xu, Yinghao and Zhao, Xinlin and Wang, Song and Xiao, Guang-Yao and Lin, Zikun and Li, Xue and Liu, Hao-Bin and Han, Henggeng and Chen, Weiyi and Weng, Yucong and Sun, Meng and Yang, Xiaohong and Liu, Jifeng},
  title = {Unveiling the Nature of G6096: A Likely Hierarchical Triple System},
  journal = {\apjl},
  year = {2026},
  volume = {1006},
  number = {2},
  pages = {L59},
  doi = {10.3847/2041-8213/ae8ae9},
  url = {https://doi.org/10.3847/2041-8213/ae8ae9},
}

@article{Shahaf2019,
  author = {Shahaf, S and Mazeh, T and Faigler, S and Holl, B},
  title = {Triage of astrometric binaries -- how to find triple systems and dormant black hole secondaries in the Gaia orbits},
  journal = {\mnras},
  year = {2019},
  volume = {487},
  number = {4},
  pages = {5610--5617},
  doi = {10.1093/mnras/stz1636},
  url = {https://doi.org/10.1093/mnras/stz1636},
  eprint = {https://academic.oup.com/mnras/article-pdf/487/4/5610/28897875/stz1636.pdf},
}

@article{Czavalinga2023,
  author = {{Czavalinga}, D.~R. and {Mitnyan}, T. and {Rappaport}, S.~A. and {Borkovits}, T. and {Gagliano}, R. and {Omohundro}, M. and {Kristiansen}, M.~H.~K. and {P{\'a}l}, A.},
  title = {{New compact hierarchical triple system candidates identified using Gaia DR3}},
  journal = {\aap},
  year = {2023},
  volume = {670},
  pages = {A75},
  doi = {10.1051/0004-6361/202245300},
  url = {https://doi.org/10.1051/0004-6361/202245300},
  adsurl = {https://ui.adsabs.harvard.edu/abs/2023A&A...670A..75C},
  archiveprefix = {arXiv},
  eprint = {2212.02903},
  primaryclass = {astro-ph.SR},
}

@article{Czavalinga2023b,
  author = {{Czavalinga}, D.~R. and {Borkovits}, T. and {Mitnyan}, T. and {Rappaport}, S.~A. and {P{\'a}l}, A.},
  title = {Four new compact triply eclipsing triples found with Gaia and TESS},
  journal = {\mnras},
  year = {2023},
  volume = {526},
  number = {2},
  pages = {2830--2845},
  doi = {10.1093/mnras/stad2759},
  url = {https://doi.org/10.1093/mnras/stad2759},
  eprint = {https://academic.oup.com/mnras/article-pdf/526/2/2830/54481412/stad2759.pdf},
}

@article{Mitnyan2024,
  author = {{Mitnyan}, T. and {Borkovits}, T. and {Czavalinga}, D.~R. and {Rappaport}, S.~A. and {P{\'a}l}, A. and {Powell}, B.~P. and {Hajdu}, T.},
  title = {{Eclipse-timing study of new hierarchical triple star candidates in the northern continuous viewing zone of TESS}},
  journal = {\aap},
  year = {2024},
  volume = {685},
  pages = {A43},
  doi = {10.1051/0004-6361/202348909},
  url = {https://doi.org/10.1051/0004-6361/202348909},
  adsurl = {https://ui.adsabs.harvard.edu/abs/2024A&A...685A..43M},
  archiveprefix = {arXiv},
  eprint = {2402.01486},
  primaryclass = {astro-ph.SR},
}

@article{Tanikawa2026,
  author = {Tanikawa, Ataru and Tajitsu, Akito and Honda, Satoshi and Maehara, Hiroyuki and Sato, Bun'ei and Masuda, Kento and Omiya, Masashi and Izumiura, Hideyuki},
  title = {Discovery of a compact hierarchical triple main-sequence star system while searching for binary stars with compact objects},
  journal = {\pasj},
  year = {2026},
  volume = {78},
  number = {3},
  pages = {1046--1058},
  doi = {10.1093/pasj/psag041},
  url = {https://doi.org/10.1093/pasj/psag041},
  eprint = {https://academic.oup.com/pasj/article-pdf/78/3/1046/68133847/psag041.pdf},
}

@article{Kovalev2024,
  author = {Kovalev, Mikhail and Zhou, Zenghua and Chen, Xuefei and Han, Zhanwen},
  title = {Detection of 12 426 SB2 candidates in the LAMOST-MRS, using a binary spectral model},
  journal = {\mnras},
  year = {2024},
  volume = {527},
  number = {1},
  pages = {521--530},
  doi = {10.1093/mnras/stad3222},
  url = {https://doi.org/10.1093/mnras/stad3222},
  eprint = {https://academic.oup.com/mnras/article-pdf/527/1/521/52762813/stad3222.pdf},
}

@ARTICLE{Duan2026,
  author = {{Duan}, X.-W. and {Chen}, Q. and {Feng}, F. and {Xiao}, G.-Y. and {Rui}, Y. and {Jones}, Hugh R.~A. and {Wang}, W.},
  title = {Binary companions from joint LAMOST radial velocities and Gaia astrometry. I. 3D orbits, true masses, and the dynamical dark--luminous divide},
  journal = {submitted to A\&A},
  year = 2026,
  note = {submitted}
}

@article{Choi2016,
  author = {Choi, Jieun and Dotter, Aaron and Conroy, Charlie and Cantiello, Matteo and Paxton, Bill and Johnson, Benjamin D.},
  title = {MESA ISOCHRONES AND STELLAR TRACKS (MIST). I. SOLAR-SCALED MODELS},
  journal = {\apj},
  year = {2016},
  volume = {823},
  number = {2},
  pages = {102},
  doi = {10.3847/0004-637X/823/2/102},
  url = {https://doi.org/10.3847/0004-637X/823/2/102},
}

@article{Dotter2016,
  author = {Dotter, Aaron},
  title = {MESA ISOCHRONES AND STELLAR TRACKS (MIST) 0: METHODS FOR THE CONSTRUCTION OF STELLAR ISOCHRONES},
  journal = {\apjs},
  year = {2016},
  volume = {222},
  number = {1},
  pages = {8},
  doi = {10.3847/0067-0049/222/1/8},
  url = {https://doi.org/10.3847/0067-0049/222/1/8},
}

@article{Xiao2026,
  author = {Xiao, Guang-Yao and Feng, Fabo and Wang, Song and Li, Kai and Rui, Yicheng and Duan, Xiao-Wei},
  title = {Detection of Dark Companions via The Combination of Eclipse Timing Variation, Hipparcos, and/or Gaia Astrometry: The Cases of V Puppis and CY Ari},
  journal = {\apj},
  year = {2026},
  volume = {998},
  number = {1},
  pages = {155},
  doi = {10.3847/1538-4357/ae314c},
  url = {https://doi.org/10.3847/1538-4357/ae314c},
}

@article{Wright2010,
  author = {Wright, Edward L. and Eisenhardt, Peter R. M. and Mainzer, Amy K. and Ressler, Michael E. and Cutri, Roc M. and Jarrett, Thomas and Kirkpatrick, J. Davy and Padgett, Deborah and McMillan, Robert S. and Skrutskie, Michael and Stanford, S. A. and Cohen, Martin and Walker, Russell G. and Mather, John C. and Leisawitz, David and Gautier, Thomas N. and McLean, Ian and Benford, Dominic and Lonsdale, Carol J. and Blain, Andrew and Mendez, Bryan and Irace, William R. and Duval, Valerie and Liu, Fengchuan and Royer, Don and Heinrichsen, Ingolf and Howard, Joan and Shannon, Mark and Kendall, Martha and Walsh, Amy L. and Larsen, Mark and Cardon, Joel G. and Schick, Scott and Schwalm, Mark and Abid, Mohamed and Fabinsky, Beth and Naes, Larry and Tsai, Chao-Wei},
  title = {THE WIDE-FIELD INFRARED SURVEY EXPLORER (WISE): MISSION DESCRIPTION AND INITIAL ON-ORBIT PERFORMANCE},
  journal = {\aj},
  year = {2010},
  volume = {140},
  number = {6},
  pages = {1868},
  doi = {10.1088/0004-6256/140/6/1868},
  url = {https://doi.org/10.1088/0004-6256/140/6/1868},
}

@article{Liu2015,
  author = {Liu, Bin and Muñoz, Diego J. and Lai, Dong},
  title = {Suppression of extreme orbital evolution in triple systems with short-range forces},
  journal = {\mnras},
  year = {2015},
  volume = {447},
  number = {1},
  pages = {747--764},
  doi = {10.1093/mnras/stu2396},
  url = {https://doi.org/10.1093/mnras/stu2396},
  eprint = {https://academic.oup.com/mnras/article-pdf/447/1/747/4907993/stu2396.pdf},
}

@article{Soderhjelm1975,
  author = {S{\"o}derhjelm, S.},
  title = {The three-body problem and eclipsing binaries. Application to Algol and lambda Tauri},
  journal = {\aap},
  year = {1975},
  volume = {42},
  pages = {229--236},
  url = {https://ui.adsabs.harvard.edu/abs/1975A&A....42..229S},
  adsurl = {https://ui.adsabs.harvard.edu/abs/1975A&A....42..229S},
}

@article{Torres2001,
  author = {Torres, Guillermo},
  title = {The Change in the Inclination Angle of the Noneclipsing Binary SS Lacertae: Future Eclipses},
  journal = {\aj},
  year = {2001},
  volume = {121},
  number = {4},
  pages = {2227--2238},
  doi = {10.1086/319942},
  url = {https://doi.org/10.1086/319942},
}

@article{Anderson2017,
  author = {Anderson, Kassandra R. and Lai, Dong and Storch, Natalia I.},
  title = {Eccentricity and spin-orbit misalignment in short-period stellar binaries as a signpost of hidden tertiary companions},
  journal = {\mnras},
  year = {2017},
  volume = {467},
  number = {3},
  pages = {3066--3082},
  doi = {10.1093/mnras/stx293},
  url = {https://doi.org/10.1093/mnras/stx293},
  eprint = {https://academic.oup.com/mnras/article-pdf/467/3/3066/10875868/stx293.pdf},
}

@article{Mazeh1976,
  author = {{Mazeh}, T. and {Shaham}, J.},
  title = {Precession of the Nodes in Some Triple Stellar Systems},
  journal = {\apjl},
  year = {1976},
  volume = {205},
  pages = {L147--L150},
  doi = {10.1086/182110},
  url = {https://doi.org/10.1086/182110},
}

@article{Carter2011,
  author = {{Carter}, Joshua A. and {Fabrycky}, Daniel C. and {Ragozzine}, Darin and {Holman}, Matthew J. and {Quinn}, Samuel N. and {Latham}, David W. and {Buchhave}, Lars A. and {Van Cleve}, Jeffrey and {Cochran}, William D. and {Cote}, Miles T. and {Endl}, Michael and {Ford}, Eric B. and {Haas}, Michael R. and {Jenkins}, Jon M. and {Koch}, David G. and {Li}, Jie and {Lissauer}, Jack J. and {MacQueen}, Phillip J. and {Middour}, Christopher K. and {Orosz}, Jerome A. and {Rowe}, Jason F. and {Steffen}, Jason H. and {Welsh}, William F.},
  title = {{KOI-126: A Triply Eclipsing Hierarchical Triple with Two Low-Mass Stars}},
  journal = {Science},
  year = {2011},
  volume = {331},
  number = {6017},
  pages = {562},
  doi = {10.1126/science.1201274},
  archivePrefix = {arXiv},
  eprint = {1102.0562},
  primaryClass = {astro-ph.SR},
  adsurl = {https://ui.adsabs.harvard.edu/abs/2011Sci...331..562C},
}

@article{Zasche2023,
  author = {Zasche, Petr and Vokrouhlick{\'y}, David and Barlow, Brad N. and Ma{\v{s}}ek, Martin},
  title = {{V907 Sco Switched to the Eclipsing Mode Again}},
  journal = {\aj},
  year = {2023},
  volume = {165},
  number = {3},
  pages = {81},
  doi = {10.3847/1538-3881/acae95},
}

@article{Vokrouhlicky2022,
  author = {{Vokrouhlick{\'y}}, David and {Zasche}, Petr},
  title = {{A New Look at the HS Hydrae System}},
  journal = {\aj},
  year = {2022},
  volume = {163},
  number = {2},
  pages = {94},
  doi = {10.3847/1538-3881/ac4492},
  url = {https://doi.org/10.3847/1538-3881/ac4492},
}

@article{Tokovinin2017,
  author = {Tokovinin, Andrei},
  title = {{Orbit Alignment in Triple Stars}},
  journal = {\apj},
  year = {2017},
  volume = {844},
  number = {2},
  pages = {103},
  doi = {10.3847/1538-4357/aa7746},
}

@article{Skrutskie2006,
  author = {Skrutskie, M. F. and Cutri, R. M. and Stiening, R. and Weinberg, M. D. and Schneider, S. and Carpenter, J. M. and Beichman, C. and Capps, R. and Chester, T. and Elias, J. and Huchra, J. and Liebert, J. and Lonsdale, C. and Monet, D. G. and Price, S. and Seitzer, P. and Jarrett, T. and Kirkpatrick, J. D. and Gizis, J. E. and Howard, E. and Evans, T. and Fowler, J. and Fullmer, L. and Hurt, R. and Light, R. and Kopan, E. L. and Marsh, K. A. and McCallon, H. L. and Tam, R. and Van Dyk, S. and Wheelock, S.},
  title = {{The Two Micron All Sky Survey (2MASS)}},
  journal = {\aj},
  year = {2006},
  volume = {131},
  number = {2},
  pages = {1163},
  doi = {10.1086/498708},
}

@article{Henden2016,
  author = {Henden, A. A. and Templeton, M. and Terrell, D. and Smith, T. C. and Levine, S. and Welch, D.},
  title = {VizieR Online Data Catalog: AAVSO Photometric All Sky Survey (APASS) DR9},
  journal = {VizieR Online Data Catalog},
  year = {2016},
  pages = {II/336},
}

@article{Astropy2013,
  author = {{Astropy Collaboration} and {Robitaille}, Thomas P. and {Tollerud}, Erik J. and {Greenfield}, Perry and {Droettboom}, Michael and {Bray}, Erik and {Aldcroft}, Tom and {Davis}, Matt and {Ginsburg}, Adam and {Price-Whelan}, Adrian M. and {Kerzendorf}, Wolfgang E. and {Conley}, Alexander and {Crighton}, Neil and {Barbary}, Kyle and {Muna}, Demitri and {Ferguson}, Henry and {Grollier}, Fr{\'e}d{\'e}ric and {Parikh}, Madhura M. and {Nair}, Prasanth H. and {Unther}, Hans M. and {Deil}, Christoph and {Woillez}, Julien and {Conseil}, Simon and {Kramer}, Roban and {Turner}, James E.~H. and {Singer}, Leo and {Fox}, Ryan and {Weaver}, Benjamin A. and {Zabalza}, Victor and {Edwards}, Zachary I. and {Azalee Bostroem}, K. and {Burke}, D.~J. and {Casey}, Andrew R. and {Crawford}, Steven M. and {Dencheva}, Nadia and {Ely}, Justin and {Jenness}, Tim and {Labrie}, Kathleen and {Lim}, Pey Lian and {Pierfederici}, Francesco and {Pontzen}, Andrew and {Ptak}, Andy and {Refsdal}, Brian and {Servillat}, Mathieu and {Streicher}, Ole},
  title = {{Astropy: A community Python package for astronomy}},
  journal = {\aap},
  year = {2013},
  volume = {558},
  pages = {A33},
  doi = {10.1051/0004-6361/201322068},
  archivePrefix = {arXiv},
  eprint = {1307.6212},
  primaryClass = {astro-ph.IM},
  adsurl = {https://ui.adsabs.harvard.edu/abs/2013A&A...558A..33A},
}

@article{Astropy2018,
  author = {{Astropy Collaboration} and Price-Whelan, A. M. and Sip{\H{o}}cz, B. M. and G{\"u}nther, H. M. and Lim, P. L. and Crawford, S. M. and Conseil, S. and Shupe, D. L. and Craig, M. W. and Dencheva, N. and Ginsburg, A. and VanderPlas, J. T. and Bradley, L. D. and P{\'e}rez-Su{\'a}rez, D. and de Val-Borro, M. and (Primary Paper Contributors) and Aldcroft, T. L. and Cruz, K. L. and Robitaille, T. P. and Tollerud, E. J. and (Astropy Coordination Committee) and Ardelean, C. and Babej, T. and Bach, Y. P. and Bachetti, M. and Bakanov, A. V. and Bamford, S. P. and Barentsen, G. and Barmby, P. and Baumbach, A. and Berry, K. L. and Biscani, F. and Boquien, M. and Bostroem, K. A. and Bouma, L. G. and Brammer, G. B. and Bray, E. M. and Breytenbach, H. and Buddelmeijer, H. and Burke, D. J. and Calderone, G. and Rodr{\'\i}guez, J. L. Cano and Cara, M. and Cardoso, J. V. M. and Cheedella, S. and Copin, Y. and Corrales, L. and Crichton, D. and D'Avella, D. and Deil, C. and Depagne, {\'E}. and Dietrich, J. P. and Donath, A. and Droettboom, M. and Earl, N. and Erben, T. and Fabbro, S. and Ferreira, L. A. and Finethy, T. and Fox, R. T. and Garrison, L. H. and Gibbons, S. L. J. and Goldstein, D. A. and Gommers, R. and Greco, J. P. and Greenfield, P. and Groener, A. M. and Grollier, F. and Hagen, A. and Hirst, P. and Homeier, D. and Horton, A. J. and Hosseinzadeh, G. and Hu, L. and Hunkeler, J. S. and Ivezi{\'c}, {\v{Z}}. and Jain, A. and Jenness, T. and Kanarek, G. and Kendrew, S. and Kern, N. S. and Kerzendorf, W. E. and Khvalko, A. and King, J. and Kirkby, D. and Kulkarni, A. M. and Kumar, A. and Lee, A. and Lenz, D. and Littlefair, S. P. and Ma, Z. and Macleod, D. M. and Mastropietro, M. and McCully, C. and Montagnac, S. and Morris, B. M. and Mueller, M. and Mumford, S. J. and Muna, D. and Murphy, N. A. and Nelson, S. and Nguyen, G. H. and Ninan, J. P. and N{\"o}the, M. and Ogaz, S. and Oh, S. and Parejko, J. K. and Parley, N. and Pascual, S. and Patil, R. and Patil, A. A. and Plunkett, A. L. and Prochaska, J. X. and Rastogi, T. and Janga, V. Reddy and Sabater, J. and Sakurikar, P. and Seifert, M. and Sherbert, L. E. and Sherwood-Taylor, H. and Shih, A. Y. and Sick, J. and Silbiger, M. T. and Singanamalla, S. and Singer, L. P. and Sladen, P. H. and Sooley, K. A. and Sornarajah, S. and Streicher, O. and Teuben, P. and Thomas, S. W. and Tremblay, G. R. and Turner, J. E. H. and Terr{\'o}n, V. and Kerkwijk, M. H. van and de la Vega, A. and Watkins, L. L. and Weaver, B. A. and Whitmore, J. B. and Woillez, J. and Zabalza, V. and (Astropy Contributors)},
  title = {{The Astropy Project: Building an Open-science Project and Status of the v2.0 Core Package}},
  journal = {\aj},
  year = {2018},
  volume = {156},
  number = {3},
  pages = {123},
  doi = {10.3847/1538-3881/aabc4f},
}

@article{Astropy2022,
  author = {{Astropy Collaboration} and Price-Whelan, Adrian M. and Lim, Pey Lian and Earl, Nicholas and Starkman, Nathaniel and Bradley, Larry and Shupe, David L. and Patil, Aarya A. and Corrales, Lia and Brasseur, C. E. and N{\"o}the, Maximilian and Donath, Axel and Tollerud, Erik and Morris, Brett M. and Ginsburg, Adam and Vaher, Eero and Weaver, Benjamin A. and Tocknell, James and Jamieson, William and van Kerkwijk, Marten H. and Robitaille, Thomas P. and Merry, Bruce and Bachetti, Matteo and G{\"u}nther, H. Moritz and Paper Authors and Aldcroft, Thomas L. and Alvarado-Montes, Jaime A. and Archibald, Anne M. and B{\'o}di, Attila and Bapat, Shreyas and Barentsen, Geert and Baz{\'a}n, Juanjo and Biswas, Manish and Boquien, M{\'e}d{\'e}ric and Burke, D. J. and Cara, Daria and Cara, Mihai and Conroy, Kyle E and Conseil, Simon and Craig, Matthew W. and Cross, Robert M. and Cruz, Kelle L. and D'Eugenio, Francesco and Dencheva, Nadia and Devillepoix, Hadrien A. R. and Dietrich, J{\"o}rg P. and Eigenbrot, Arthur Davis and Erben, Thomas and Ferreira, Leonardo and Foreman-Mackey, Daniel and Fox, Ryan and Freij, Nabil and Garg, Suyog and Geda, Robel and Glattly, Lauren and Gondhalekar, Yash and Gordon, Karl D. and Grant, David and Greenfield, Perry and Groener, Austen M. and Guest, Steve and Gurovich, Sebastian and Handberg, Rasmus and Hart, Akeem and Hatfield-Dodds, Zac and Homeier, Derek and Hosseinzadeh, Griffin and Jenness, Tim and Jones, Craig K. and Joseph, Prajwel and Kalmbach, J. Bryce and Karamehmetoglu, Emir and Ka{\l}uszy{\'n}ski, Miko{\l}aj and Kelley, Michael S. P. and Kern, Nicholas and Kerzendorf, Wolfgang E. and Koch, Eric W. and Kulumani, Shankar and Lee, Antony and Ly, Chun and Ma, Zhiyuan and MacBride, Conor and Maljaars, Jakob M. and Muna, Demitri and Murphy, N. A. and Norman, Henrik and O'Steen, Richard and Oman, Kyle A. and Pacifici, Camilla and Pascual, Sergio and Pascual-Granado, J. and Patil, Rohit R. and Perren, Gabriel I and Pickering, Timothy E. and Rastogi, Tanuj and Roulston, Benjamin R. and Ryan, Daniel F and Rykoff, Eli S. and Sabater, Jose and Sakurikar, Parikshit and Salgado, Jes{\'u}s and Sanghi, Aniket and Saunders, Nicholas and Savchenko, Volodymyr and Schwardt, Ludwig and Seifert-Eckert, Michael and Shih, Albert Y. and Jain, Anany Shrey and Shukla, Gyanendra and Sick, Jonathan and Simpson, Chris and Singanamalla, Sudheesh and Singer, Leo P. and Singhal, Jaladh and Sinha, Manodeep and Sip{\H{o}}cz, Brigitta M. and Spitler, Lee R. and Stansby, David and Streicher, Ole and {\v{S}}umak, Jani and Swinbank, John D. and Taranu, Dan S. and Tewary, Nikita and Tremblay, Grant R. and Val-Borro, Miguel de and Van Kooten, Samuel J. and Vasovi{\'c}, Zlatan and Verma, Shresth and de Miranda Cardoso, Jos{\'e} Vin{\'\i}cius and Williams, Peter K. G. and Wilson, Tom J. and Winkel, Benjamin and Wood-Vasey, W. M. and Xue, Rui and Yoachim, Peter and Zhang, Chen and Zonca, Andrea and Astropy Project Contributors},
  title = {{The Astropy Project: Sustaining and Growing a Community-oriented Open-source Project and the Latest Major Release (v5.0) of the Core Package}},
  journal = {\apj},
  year = {2022},
  volume = {935},
  number = {2},
  pages = {167},
  doi = {10.3847/1538-4357/ac7c74},
}

@article{Harris2020,
  author = {{Harris}, Charles R. and {Millman}, K. Jarrod and {van der Walt}, St{\'e}fan J. and {Gommers}, Ralf and {Virtanen}, Pauli and {Cournapeau}, David and {Wieser}, Eric and {Taylor}, Julian and {Berg}, Sebastian and {Smith}, Nathaniel J. and {Kern}, Robert and {Picus}, Matti and {Hoyer}, Stephan and {van Kerkwijk}, Marten H. and {Brett}, Matthew and {Haldane}, Allan and {del R{\'\i}o}, Jaime Fern{\'a}ndez and {Wiebe}, Mark and {Peterson}, Pearu and {G{\'e}rard-Marchant}, Pierre and {Sheppard}, Kevin and {Reddy}, Tyler and {Weckesser}, Warren and {Abbasi}, Hameer and {Gohlke}, Christoph and {Oliphant}, Travis E.},
  title = {{Array programming with NumPy}},
  journal = {Nature},
  year = {2020},
  volume = {585},
  number = {7825},
  pages = {357--362},
  doi = {10.1038/s41586-020-2649-2},
  archivePrefix = {arXiv},
  eprint = {2006.10256},
  primaryClass = {cs.MS},
  adsurl = {https://ui.adsabs.harvard.edu/abs/2020Natur.585..357H},
}

@article{Virtanen2020,
  author = {{Virtanen}, Pauli and {Gommers}, Ralf and {Oliphant}, Travis E. and {Haberland}, Matt and {Reddy}, Tyler and {Cournapeau}, David and {Burovski}, Evgeni and {Peterson}, Pearu and {Weckesser}, Warren and {Bright}, Jonathan and others},
  title = {{SciPy 1.0: fundamental algorithms for scientific computing in Python}},
  journal = {NatMe},
  year = {2020},
  volume = {17},
  pages = {261--272},
  doi = {10.1038/s41592-019-0686-2},
  archivePrefix = {arXiv},
  eprint = {1907.10121},
  primaryClass = {cs.MS},
  adsurl = {https://ui.adsabs.harvard.edu/abs/2020NaMet..17..261V},
}

@article{Hunter2007,
  author = {Hunter, John D.},
  title = {{Matplotlib: A 2D Graphics Environment}},
  journal = {CSE},
  year = {2007},
  volume = {9},
  number = {3},
  pages = {90--95},
  doi = {10.1109/MCSE.2007.55},
}

@article{Pollacco2006,
  author = {{Pollacco}, D.~L. and {Skillen}, I. and {Collier Cameron}, A. and {Christian}, D.~J. and {Hellier}, C. and {Irwin}, J. and {Lister}, T.~A. and {Street}, R.~A. and {West}, R.~G. and {Anderson}, D.~R. and others},
  title = {{The WASP Project and the SuperWASP Cameras}},
  journal = {\pasp},
  year = {2006},
  volume = {118},
  number = {848},
  pages = {1407--1418},
  doi = {10.1086/508556},
  archivePrefix = {arXiv},
  eprint = {astro-ph/0608454},
  primaryClass = {astro-ph},
  adsurl = {https://ui.adsabs.harvard.edu/abs/2006PASP..118.1407P},
}

@article{Butters2010,
  author = {{Butters}, O.~W. and {West}, R.~G. and {Anderson}, D.~R. and {Collier Cameron}, A. and {Clarkson}, W.~I. and {Enoch}, B. and {Haswell}, C.~A. and {Hellier}, C. and {Horne}, K. and {Joshi}, Y. and {Kane}, S.~R. and {Lister}, T.~A. and {Maxted}, P.~F.~L. and {Parley}, N. and {Pollacco}, D. and {Smalley}, B. and {Street}, R.~A. and {Todd}, I. and {Wheatley}, P.~J. and {Wilson}, D.~M.},
  title = {{The first WASP public data release}},
  journal = {\aap},
  year = {2010},
  volume = {520},
  pages = {L10},
  doi = {10.1051/0004-6361/201015655},
  archivePrefix = {arXiv},
  eprint = {1009.5306},
  primaryClass = {astro-ph.EP},
  adsurl = {https://ui.adsabs.harvard.edu/abs/2010A&A...520L..10B},
}

@article{Pepper2007,
  author = {{Pepper}, Joshua and {Pogge}, Richard W. and {DePoy}, D.~L. and {Marshall}, J.~L. and {Stanek}, K.~Z. and {Stutz}, Amelia M. and {Poindexter}, Shawn and {Siverd}, Robert and {O'Brien}, Thomas P. and {Trueblood}, Mark and others},
  title = {{The Kilodegree Extremely Little Telescope (KELT): A Small Robotic Telescope for Large-Area Synoptic Surveys}},
  journal = {\pasp},
  year = {2007},
  volume = {119},
  number = {858},
  pages = {923--935},
  doi = {10.1086/521836},
  archivePrefix = {arXiv},
  eprint = {0704.0460},
  primaryClass = {astro-ph},
  adsurl = {https://ui.adsabs.harvard.edu/abs/2007PASP..119..923P},
}

@article{Shappee2014,
  author = {Shappee, B. J. and Prieto, J. L. and Grupe, D. and Kochanek, C. S. and Stanek, K. Z. and others},
  title = {The Man behind the Curtain: X-Rays Drive the UV through NIR Variability in the 2013 Active Galactic Nucleus Outburst in NGC 2617},
  journal = {\apj},
  year = {2014},
  volume = {788},
  pages = {48},
  doi = {10.1088/0004-637X/788/1/48},
}

@article{Jayasinghe2018,
  author = {{Jayasinghe}, T. and {Kochanek}, C.~S. and {Stanek}, K.~Z. and {Shappee}, B.~J. and {Holoien}, T.~W.-S. and {Thompson}, Todd A. and {Prieto}, J.~L. and {Dong}, Subo and {Pawlak}, M. and {Shields}, J.~V. and others},
  title = {{The ASAS-SN catalogue of variable stars I: The Serendipitous Survey}},
  journal = {\mnras},
  year = {2018},
  volume = {477},
  number = {3},
  pages = {3145--3163},
  doi = {10.1093/mnras/sty838},
  archivePrefix = {arXiv},
  eprint = {1803.01001},
  primaryClass = {astro-ph.SR},
  adsurl = {https://ui.adsabs.harvard.edu/abs/2018MNRAS.477.3145J},
}

@article{Kochanek2017,
  author = {Kochanek, C. S. and Shappee, B. J. and Stanek, K. Z. and Holoien, T. W.-S. and Thompson, Todd A. and others},
  title = {The All-Sky Automated Survey for Supernovae (ASAS-SN) Light Curve Server v1.0},
  journal = {\pasp},
  year = {2017},
  volume = {129},
  number = {980},
  pages = {104502},
  doi = {10.1088/1538-3873/aa80d9},
}

@article{Pecaut2013,
  author = {Pecaut, Mark J. and Mamajek, Eric E.},
  title = {Intrinsic Colors, Temperatures, and Bolometric Corrections of Pre-main-sequence Stars},
  journal = {\apjs},
  year = {2013},
  volume = {208},
  number = {1},
  pages = {9},
  doi = {10.1088/0067-0049/208/1/9},
}

@article{ElBadry2026,
  author = {{El-Badry}, Kareem and {Latham}, David W. and {Rix}, Hans-Walter and {Bieryla}, Allyson and {Buchhave}, Lars A. and {Shahaf}, Sahar and {Mazeh}, Tsevi and {M{\"u}ller-Horn}, Johanna and {Nagarajan}, Pranav and {Yamaguchi}, Natsuko and others},
  title = {{Spectroscopic follow-up of compact object binary candidates from Gaia DR3: White dwarfs, neutron stars, black holes, and the parallax zeropoint}},
  journal = {The Open Journal of Astrophysics},
  year = {2026},
  volume = {9},
  doi = {10.33232/001c.172274},
  archivePrefix = {arXiv},
  eprint = {2608.06453},
  primaryClass = {astro-ph.SR},
  adsurl = {https://ui.adsabs.harvard.edu/abs/2026OJAp....972274E},
}

@article{Luo2016,
  author = {Luo, Liantong and Katz, Boaz and Dong, Subo},
  title = {Double-averaging can fail to characterize the long-term evolution of Lidov-Kozai Cycles and derivation of an analytical correction},
  journal = {\mnras},
  year = {2016},
  volume = {458},
  number = {3},
  pages = {3060--3074},
  doi = {10.1093/mnras/stw475},
  url = {https://doi.org/10.1093/mnras/stw475},
}

@article{Grishin2018,
  author = {Grishin, Evgeni and Perets, Hagai B. and Fragione, Giacomo},
  title = {Quasi-secular evolution of mildly hierarchical triple systems: analytics and applications for GW sources and hot Jupiters},
  journal = {\mnras},
  year = {2018},
  volume = {481},
  number = {4},
  pages = {4907--4923},
  doi = {10.1093/mnras/sty2477},
  url = {https://doi.org/10.1093/mnras/sty2477},
}

@article{Tamayo2020,
  author = {{Tamayo}, Daniel and {Rein}, Hanno and {Shi}, Pengshuai and {Hernandez}, David M.},
  title = {{REBOUNDx: a library for adding conservative and dissipative forces to otherwise symplectic N-body integrations}},
  journal = {\mnras},
  year = {2020},
  volume = {491},
  number = {2},
  pages = {2885--2901},
  doi = {10.1093/mnras/stz2870},
  archivePrefix = {arXiv},
  eprint = {1908.05634},
  primaryClass = {astro-ph.EP},
  adsurl = {https://ui.adsabs.harvard.edu/abs/2020MNRAS.491.2885T},
}

@article{Lu2023,
  author = {{Lu}, Tiger and {Rein}, Hanno and {Tamayo}, Daniel and {Hadden}, Sam and {Mardling}, Rosemary and {Millholland}, Sarah C. and {Laughlin}, Gregory},
  title = {{Self-consistent Spin, Tidal, and Dynamical Equations of Motion in the REBOUNDx Framework}},
  journal = {\apj},
  year = {2023},
  volume = {948},
  number = {1},
  pages = {41},
  doi = {10.3847/1538-4357/acc06d},
  archivePrefix = {arXiv},
  eprint = {2303.00006},
  primaryClass = {astro-ph.EP},
  adsurl = {https://ui.adsabs.harvard.edu/abs/2023ApJ...948...41L},
}

@article{Fabrycky2007,
  author = {Fabrycky, Daniel and Tremaine, Scott},
  title = {Shrinking Binary and Planetary Orbits by Kozai Cycles with Tidal Friction},
  journal = {\apj},
  year = {2007},
  volume = {669},
  number = {2},
  pages = {1298--1315},
  doi = {10.1086/521702},
  url = {https://doi.org/10.1086/521702},
}

@article{Moe2018,
  author = {{Moe}, Maxwell and {Kratter}, Kaitlin M.},
  title = {{Dynamical Formation of Close Binaries during the Pre-main-sequence Phase}},
  journal = {\apj},
  year = {2018},
  volume = {854},
  number = {1},
  pages = {44},
  doi = {10.3847/1538-4357/aaa6d2},
  archivePrefix = {arXiv},
  eprint = {1706.09894},
  primaryClass = {astro-ph.SR},
  adsurl = {https://ui.adsabs.harvard.edu/abs/2018ApJ...854...44M},
}

@article{Yee2026,
  author = {Yee, Samuel W. and Petigura, Erik A. and Brahm, Rafael and Collins, Karen A. and Srdoc, Gregor and Latham, David W. and Bieryla, Allyson and DiTomasso, Victoria and Henning, Thomas and Jord{\'a}n, Andr{\'e}s and Trifonov, Trifon and Barkaoui, Khalid and Evans, Phil and Jehin, Emmanuel and Ghachoui, Mourad and Gillon, Micha{\"e}l and Girardin, Eric and Murgas, Felipe and Palle, Enric and Popowicz, Adam and Radford, Don J. and Relles, Howard M. and Schwarz, Richard P. and Shporer, Avi and Soubkiou, Abderahmane and Watanabe, David and Wilkin, Francis P. and Ziegler, Carl},
  title = {{A TESS Disappearing Act: The Vanishing Transits of TOI-2535 b}},
  journal = {\apjl},
  year = {2026},
  volume = {1011},
  number = {2},
  pages = {L60},
  doi = {10.3847/2041-8213/aeb4a4},
}

@article{Xiao2026b,
  author = {Xiao, Guang-Yao and Teng, Huan-Yu and Huang, Xiumin and Feng, Fabo and Lai, Dong and Dai, Fei and Liu, Yu-Juan},
  title = {{Transiting Planetary Systems with Distant Giant Companions Remain Moderately Coplanar}},
  journal = {\apjl},
  year = {2026},
  volume = {1009},
  number = {1},
  pages = {L6},
  doi = {10.3847/2041-8213/aea15e},
}

@article{Vousden2016,
  author = {{Vousden}, W.~D. and {Farr}, W.~M. and {Mandel}, I.},
  title = {{Dynamic temperature selection for parallel tempering in Markov chain Monte Carlo simulations}},
  journal = {\mnras},
  year = {2016},
  volume = {455},
  number = {2},
  pages = {1919--1937},
  doi = {10.1093/mnras/stv2422},
  archivePrefix = {arXiv},
  eprint = {1501.05823},
  primaryClass = {astro-ph.IM},
  adsurl = {https://ui.adsabs.harvard.edu/abs/2016MNRAS.455.1919V},
}

@article{Kipping2013,
  author = {Kipping, David M.},
  title = {Efficient, uninformative sampling of limb darkening coefficients for two-parameter laws},
  journal = {\mnras},
  year = {2013},
  volume = {435},
  pages = {2152--2160},
  doi = {10.1093/mnras/stt1435},
  number = {3},
}

@article{Cardelli1989,
  author = {{Cardelli}, Jason A. and {Clayton}, Geoffrey C. and {Mathis}, John S.},
  title = {{The Relationship between Infrared, Optical, and Ultraviolet Extinction}},
  journal = {\apj},
  year = {1989},
  volume = {345},
  pages = {245},
  doi = {10.1086/167900},
  adsurl = {https://ui.adsabs.harvard.edu/abs/1989ApJ...345..245C},
}

@article{Akar2026,
  author = {{Akar}, Furkan and {Ba{\c{s}}t{\"u}rk}, {\"O}zg{\"u}r},
  title = {{A joint astrometric, eclipse timing and radial velocity orbit for the hierarchical triple V819 Herculis}},
  journal = {arXiv e-prints},
  year = {2026},
  pages = {arXiv:2611.09551},
  doi = {10.48550/arXiv.2611.09551},
  archivePrefix = {arXiv},
  eprint = {2611.09551},
  primaryClass = {astro-ph.SR},
  adsurl = {https://ui.adsabs.harvard.edu/abs/2026arXiv261109551A},
}

@article{Borkovits2019,
  author = {{Borkovits}, T. and {Rappaport}, S. and {Kaye}, T. and {Isaacson}, H. and {Vanderburg}, A. and {Howard}, A.~W. and {Kristiansen}, M.~H. and {Omohundro}, M.~R. and {Schwengeler}, H.~M. and {Terentev}, I.~A. and {Shporer}, A. and {Relles}, H. and {Villanueva}, Jr., S. and {Tan}, T.~G. and {Col{\'o}n}, K.~D. and {Blex}, J. and {Haas}, M. and {Cochran}, W. and {Endl}, M.},
  title = {{Photodynamical analysis of the triply eclipsing hierarchical triple system EPIC 249432662}},
  journal = {\mnras},
  year = {2019},
  volume = {483},
  number = {2},
  pages = {1934--1951},
  doi = {10.1093/mnras/sty3157},
  archivePrefix = {arXiv},
  eprint = {1809.04366},
  primaryClass = {astro-ph.SR},
  adsurl = {https://ui.adsabs.harvard.edu/abs/2019MNRAS.483.1934B},
}

@article{He2025,
  author = {He, Tongyu and Li, Jiao and Li, Jiangdan and Xiong, Jianping and Zhang, Xiaobin and Kovalev, Mikhail and Cheng, Qiyuan and Guo, Sufen and Yang, Mingkuan and Chen, Xuefei and Han, Zhanwen},
  title = {{Identifying Hierarchically Triple Star Systems with Gaia DR3 and LAMOST}},
  journal = {\aj},
  year = {2025},
  volume = {169},
  number = {2},
  pages = {60},
  doi = {10.3847/1538-3881/ad9915},
}

@article{Yenawine2022,
  author = {{Yenawine}, Mitchell E. and {Welsh}, William F. and {Orosz}, Jerome A. and {Bieryla}, Allyson and {Cochran}, William D. and {Endl}, Michael and {Latham}, David W. and {Quinn}, Samuel N. and {Short}, Donald R. and {Windmiller}, Gur},
  title = {{Photodynamical Modeling of the Fascinating Eclipses in the Triple-star System KOI-126}},
  journal = {\apj},
  year = {2022},
  volume = {924},
  number = {2},
  pages = {66},
  doi = {10.3847/1538-4357/ac31b8},
  archivePrefix = {arXiv},
  eprint = {2110.09661},
}

@article{Dotter2026,
  author = {Dotter, A. and Bauer, E. and Park, M. J. and Conroy, C. and Milone, A. and Joyce, M. and Cantiello, M.},
  title = {MESA Isochrones and Stellar Tracks. II. Models with alpha-enhanced chemical composition},
  journal = {\apjs},
  year = {2026},
  volume = {283},
  pages = {64},
  doi = {10.3847/1538-4365/ae48f3},
  url = {https://doi.org/10.3847/1538-4365/ae48f3},
  eprint = {2602.22012},
  archivePrefix = {arXiv}
}

@article{terBraak2006,
  author = {{ter Braak}, Cajo J.~F.},
  title = {{A Markov Chain Monte Carlo version of the genetic algorithm Differential Evolution: easy Bayesian computing for real parameter spaces}},
  journal = {Statistics and Computing},
  year = {2006},
  volume = {16},
  number = {3},
  pages = {239--249},
  doi = {10.1007/s11222-006-8769-1},
}
\bibliographystyle{aasjournalv7}
\end{document}